\documentclass[11pt]{article}

\usepackage{epsfig,amsmath,latexsym,amssymb}
\usepackage{graphicx}
\usepackage{lscape}
\usepackage{picture, eso-pic, tikz} 
\usepackage{dsfont}
\usepackage{listings}
\usepackage{changes}
\usepackage{hyperref}
\usepackage{bbding}
\usepackage{tikz-cd}
\usepackage{mathtools}
\usepackage{microtype,booktabs,longtable,tabularx,array,needspace}

\usepackage{algorithm}
\usepackage{algpseudocode}
\usepackage{microtype}

\renewcommand{\baselinestretch}{1.1}
\def\E{{\mathbb E}}  
\def\Beweis{\footnotesize}

\newcommand{\Remm}[1]{}
\newtheorem{theo}{Theorem}[section]
\newtheorem{lemma}[theo]{Lemma}
\newtheorem{prop}[theo]{Proposition}

\newtheorem{cor}[theo]{Corollary}

\newtheorem{model ass}[theo]{Model Assumptions}
\newtheorem{ass}[theo]{Assumptions}

\def\EndProof{\hfill {\scriptsize $\Box$}}

\numberwithin{equation}{section}

\definecolor{MyGray}{rgb}{0.92,0.92,0.92}


\definecolor{British racing}{rgb}{0.0, 0.5, 0.0}

\def\bTheta{\boldsymbol{\Theta}}

\def\bgamma{\boldsymbol{\gamma}}

\def\b0{\boldsymbol{0}}

\usepackage{todonotes}
\newcommand{\Comments}{1}
\newcommand{\mynote}[2]{\ifnum\Comments=1\textcolor{#1}{#2}\fi}
\newcommand{\mytodo}[2]{\ifnum\Comments=1%
  \todo[linecolor=#1!80!black,backgroundcolor=#1,bordercolor=#1!80!black]{#2}\fi}

\ifnum\Comments=1               
\fi

\ifnum\Comments=1               
\fi

\ifnum\Comments=1               
\fi

\begin{document}

\author{Mario V.~W{\"u}thrich\footnote{\normalfont Department of Mathematics, ETH Zurich, mario.wuethrich@math.ethz.ch}}

\date{\today}
\title{{\sc Exponential Seniority Weighting\\ Chain-Ladder Reserving}}
\maketitle

\begin{abstract}
  \noindent
We develop a tractable state-space version of the chain-ladder model. The state-space dynamics acts on the accident year axis which, in contrast to the classical chain-ladder model, makes accident years dependent. This state-space version yields an exponential seniority weighting in the chain-ladder factor estimation, and it admits an analytical likelihood representation that enables us to compute the seniority weighted chain-ladder reserves and their prediction uncertainty: the one-year uncertainty is available in closed form and the total uncertainty requires a Monte Carlo simulation from the fully tractable likelihood representation.

\bigskip

\end{abstract}

\section{Introduction}

\begin{figure}[htb!]
\begin{center}
\begin{minipage}[t]{0.95\textwidth}
\begin{center}
\includegraphics[width=\textwidth]{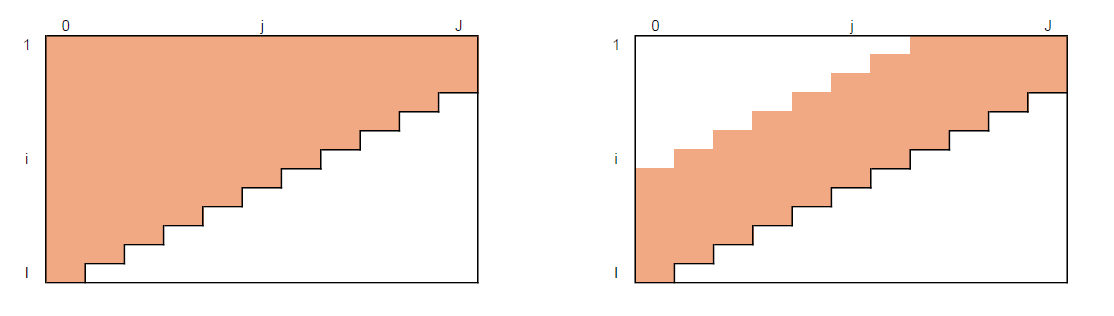}
\end{center}
\end{minipage}
\end{center}
\vspace{-.7cm}
\caption{(lhs) Available information at time $I$, (rhs) observations selected for model fitting.}
\label{Figure1}
\end{figure}

Predicting outstanding insurance claims requires both a forecast of future payments and a quantification of the uncertainty in this forecast. Mack's \cite{Mack} chain-ladder (CL) method is one of the early derivations that quantified the prediction uncertainty of the CL reserving method in terms of the conditional mean squared error of prediction (MSEP). Mack's \cite{Mack} uncertainty estimate was derived in a distribution-free stochastic version of the CL method, and still today it is the state-of-the-art in claims reserving and its risk management implications. The broader literature considers alternative distributional assumptions, estimation procedures and predictive models; see, e.g., England--Verrall \cite{EV} and W\"uthrich--Merz \cite{WM}. Within this literature, an important modeling choice is how information is shared across accident years when estimating a development pattern. The classical CL specification uses a common expected development factor for each development period, which can be seen as a mean-stationarity property. In practice, however, this mean-stationarity property is often violated, as external factors and portfolio changes may lead to trends in the development pattern. To cope with such trends, practitioners often use a rolling window estimation approach, meaning that instead of considering all past observations, they only use the more recent ones (in a given time window). This is illustrated in Figure \ref{Figure1}: the left-hand side (lhs) of that figure considers all past observations indicated by the orange trapezoid, whereas a rolling window approach only considers a recent diagonal slab, as shown on the right-hand side (rhs) of Figure \ref{Figure1}.

This rolling window approach seems problematic for two reasons. First, it neglects the observations in the upper-left part of the observed claims array. Second, the need for considering an estimation window indicates that the model assumptions are violated, thus, one works in an inappropriate stochastic reserving setting. Below, we consider a full stochastic model that results in an exponential seniority weighting approach, see  \eqref{CL factor for examples}, that is, the optimal CL factors are in our stochastic model given by
\begin{equation}\label{exponential weighting focus}
  \widehat{f}^{(I)}_j = \sum_{i=1}^{I-j-1} \alpha_{i,j}\,
  \frac{C_{i,j+1}}{C_{i,j}} \qquad \text{ with } \qquad \alpha_{i,j}=\frac{q^{I-j-1-i}\, C_{i,j}}{\sum_{k=1}^{I-j-1}q^{I-j-1-k} \,C_{k,j}}.
\end{equation}
The crucial part in this estimation formula is the exponential weighting $q^{I-j-1-i}$ for a given parameter $q\in (0,1]$. For the latest observation, we have $i=I-j-1$, thus, $q^{I-j-1-i}=1$; for the second latest observation $i=I-j-2$, thus, $q^{I-j-1-i}=q$; etc. For $q=1$, this yields the classical CL factors of Mack \cite{Mack}, and for $q<1$, the observations receive an exponential seniority weighting. A related estimation approach has been considered in Marais--Grove \cite[formula (4)]{MaraisGrove}, however, this proposal is not based on a fully-fledged stochastic reserving model, but it manually imposes this structure by formulating a corresponding estimation objective. We introduce a full stochastic model that allows us to analytically compute the predictors resulting precisely in such an exponential seniority weighting \eqref{exponential weighting focus}, i.e., this exponential estimation procedure is a natural consequence of our model assumptions. This full stochastic model provides us with the advantage of being able to explicitly simulate the conditional MSEP in a correctly specified stochastic reserving environment, and we receive explicit closed-form results for the one-year claims development uncertainty view. In particular, to implement 
a weighting approach \eqref{exponential weighting focus} in a state-space model requires dependence along the accident year axis, which is neither the case in Mack \cite{Mack2}, Marais--Grove \cite{MaraisGrove} nor in Sloma \cite{Sloma}. This literature considers weighting, but the weighting is added manually (somehow retrospectively) and these weights are not the natural product of a full stochastic model.

Furthermore, Marais--Grove \cite[Section 4.2]{MaraisGrove} add regularization with a prior development pattern. Our stochastic reserving model admits a similar regularization version because we work in an observation-driven state-space model where regularization is a natural implication from the latent state-space process. A unregluarized version is then obtained by letting the latent state-space process be non-informative; this is rather similar to the Bayesian considerations in W\"uthrich--Merz \cite[Section 2.2]{WM2015}.

Early versions of state-space models for claims reserving were considered by de Jong--Zehnwirth \cite{deJong} and
Verrall \cite{Verrall1989, Verrall1994}; a nice historical overview is given in Chukhrova--Johannssen \cite{ChukhrovaJohannssen}. In a multivariate log-normal context there is full modeling flexibility of the dependence structure along all time directions; see W\"uthrich--Merz \cite[Chapter 4]{WM2015}. Most of these state-space proposals have in common that they do not use a linear scaling in the conditional variance as proposed and introduced by Mack \cite{Mack}, thus, they do not directly lift Mack's \cite{Mack} distribution-free CL model to a version that allows for seniority weighting. 
Chukhrova--Johannssen \cite[Model Assumptions 2]{ChukhrovaJohannssen} lift Mack's \cite{Mack} CL model to a distribution-free state-space model, which is then solved by applying the Kalman filter. However, their model does not allow for a seniority weighting in the accident year direction because the state-space dynamics evolves along the development year axis, whereas our proposal will move along the accident year axis. The way we are going to process information is similar to Cipra--Hendrych \cite{CipraHendrych}, but in contrast to that reference, we process cumulative claims and we ensure that Mack's \cite{Mack} linear conditional variance assumption is fulfilled, so that Mack's \cite{Mack} method can be seen as a boundary case of our exponential seniority weighting claims reserving model.

More explicitly, we combine the gamma-gamma Bayes' CL model of W\"uthrich--Merz \cite{WM2015} with the observation-driven state-space model of Ahn et al.~\cite{Ahn2}. This yields a full stochastic model that allows us to compute all relevant quantities because we can express the likelihood of the observations in closed form in this model. More specifically, the gamma-gamma Bayes' CL model considers conjugate priors that allow for explicit computations of posteriors, and they also allow us to incorporate Mack's linear conditional variance assumption. This model can be marginalized by integrating out the latent states resulting in prime-beta distributions. We then lift this model to a state-space version using the observation-driven state-space dynamics of Ahn et al.~\cite{Ahn2}, which was inspired by the econometrics models of Smith--Miller \cite{SmithMiller} and Harvey--Fernandes \cite{Harvey}. This observation-driven state-space model retains analytical tractability, and this precisely makes it possible to derive our results.

Our contributions are the following. We introduce a full stochastic claims reserving model that is analytically tractable. This model allows us to determine the claims reserves, the conditional MSEP of the total run-off and the conditional MSEP of the one-year claims development result. This model has the nice property that it allows us to select the hyper-parameters such that it yields the CL claims reserves with an exponential seniority weighting as a natural result from the model assumptions. Its boundary (special) case are the classical CL reserves, and the resulting uncertainty formulas are lower bounded by Mack's \cite{Mack} MSEP formula and by the Merz--W\"uthrich \cite{MW2008} one-year formula; this is in complete analogy to the considerations in W\"uthrich--Merz \cite[Section 2.2]{WM2015}. To the best of our knowledge, this is the first model where all these quantities can be fully derived including an exponential seniority weighting in the CL factors, i.e., yielding consistent results within a unified stochastic model where the seniority weighting is a natural consequence of the model assumptions, and not a manually enforced one.

\medskip

{\bf Organization.} In the next section, we introduce the claims development model, and we discuss basic properties.
Section \ref{Filtering, credibility, and mean-stationarity} presents the main properties of the model -- mean-stationarity of the latent state-space process -- and it derives different versions of the CL factors.
Section \ref{Ultimate claim predictions} discusses ultimate claims predictions and the special case of exponential seniority weighting in the CL factors. These computations are feasible because we have tractable likelihoods with explicit factorizations. Sections \ref{Total prediction uncertainty} and \ref{CDR prediction uncertainty} give the results for the total prediction uncertainty and the one-year uncertainty MSEPs, respectively. Section \ref{sec: Example} gives a numerical example on a standard dataset, and in Section \ref{sec: Conclusion} we conclude.

\section{The model}
\subsection{Claims prediction and the claims reserving problem}
Let $C_{i,j}$ denote the cumulative claims in accident year $i \in \{1, \ldots, I\}$ after development year $j \in \{0,\ldots, J\}$, and assume that these cumulative claims are strictly positive. We generally assume $I>J$.
The total claim of accident year $i \in \{1, \ldots, I\}$ is given by the {\it ultimate claim} $C_{i,J}$. This ultimate claim is the main object of interest that we aim to forecast based on the available information set at time $I$, given by the upper trapezoid
\begin{equation*}
  {\cal D}_I = \sigma\left\{ C_{i,j};~i+j\le I, \, 1\le i \le I, \, 0 \le j \le J \right\}.
\end{equation*}
This information set specifies all observations in the complete rows $i\le I-J$, and the observed ones in the incomplete rows $i>I-J$, i.e., given up to the claims development index
\begin{equation}\label{index di}
  d_i = \min \left\{ J, I-i \right\},
\end{equation}
this is illustrated in Figure \ref{Figure1} (lhs).
Our goal is to compute the best-estimates of the ultimate claims at time $I$ for accident years $i+J>I$,
given by
\begin{equation*}
\E \left[C_{i,J}\middle| {\cal D}_I\right].
\end{equation*}
The accident years $i\le I-J$ do not need predictions because they are fully observed at time $I$ ($d_i=J$).
Our goal is to compute these best-estimates under given model assumptions. If the cumulative claims $C_{i,j}$ correspond to cumulative payments, then the claims reserves for the {\it outstanding loss liabilities} of
accident years $i+J>I$ are given by
\begin{equation*}
R_i = \E \left[C_{i,J}\middle| {\cal D}_I\right] -C_{i,d_i}.
\end{equation*}
The starting point to compute the best-estimates and the claims reserves
is a Bayesian version of Mack's CL model \cite{Mack}. Namely, we consider an accident-year dynamic version of the gamma-gamma Bayes CL model of W\"uthrich--Merz \cite{WM2015}. The accident-year dynamic part uses the observation-driven state-space model ideas of Ahn et al.~\cite{Ahn2} involving latent state-space processes $\Theta_{1:I,j}$ for $j\in \{0,\ldots, J-1\}$, i.e., these latent processes evolve on the accident year axis for fixed development period $j$. Before doing so, we introduce the relevant information sets.

\subsection{Information sets}

We introduce the following notation. Generically, we write for vectors
$C_{1:i,j}=(C_{1,j}, \ldots, C_{i,j})$ and $C_{i,0:j}=(C_{i,0}, \ldots, C_{i,j})$, and for 2D tensors $C_{1:i,0:j}=(C_{l,k})_{l=1, \ldots, i, \, k=0, \ldots, j}$. An empty sequence $C_{1:0,j}$ or an empty 2D tensor $C_{1:0,0:j}$ generates the trivial $\sigma$-field.

\begin{figure}[htb!]
\begin{center}
\begin{minipage}[t]{0.5\textwidth}
\begin{center}
\includegraphics[width=\textwidth]{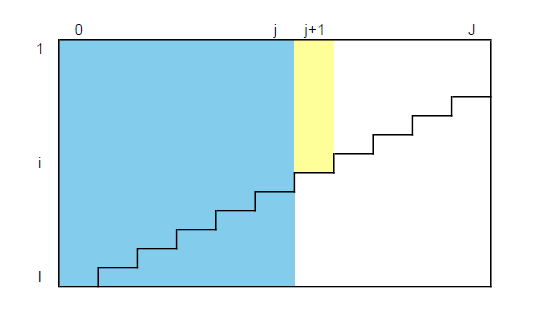}
\end{center}
\end{minipage}
\end{center}
\vspace{-.7cm}
\caption{Information set ${\cal H}_{i,j}$ with ${\cal B}_j$ in blue color and 
$\sigma( C_{1:i,j+1})$ in yellow color.}
\label{Figure2}
\end{figure}

We introduce the relevant information sets. We define
\begin{equation*}
  {\cal B}_j = \sigma \left( C_{1:I,0:j}\right) \qquad \qquad \text{ for $j \in \{0, \ldots J\}$},
\end{equation*}
this is the complete rectangle up to development period $j$, and
\begin{equation}\label{columnwise sigma algebra}
{\cal H}_{i,j} = {\cal B}_j \vee \sigma \left( C_{1:i,j+1}\right)
\qquad \text{ for $j \in \{0, \ldots J-1\}$ and $i \in \{0,\ldots, I\}$},
\end{equation}
this information set is illustrated in Figure \ref{Figure2}.
We interpret ${\cal H}_{i-1,j}$ as the history before $C_{i,j+1}$ has been observed, and ${\cal H}_{i,j}$ is the history thereafter, i.e., $C_{i,j+1}$ is ${\cal H}_{i,j}$-measurable. These $\sigma$-fields do not respect the trapezoidal structure of ${\cal D}_I$, see Figure \ref{Figure2}, but they are necessary to define our claims reserving model. These definitions yield the filtration
\begin{equation*}
\ldots \, \subseteq\,  {\cal H}_{I,j-1}={\cal B}_j={\cal H}_{0,j}\, \subseteq\, {\cal H}_{1,j}\,\subseteq\, \ldots 
\,\subseteq\, {\cal H}_{I-1,j}\,\subseteq\, {\cal H}_{I,j}={\cal B}_{j+1}= {\cal H}_{0,j+1}\,\subseteq\, \ldots.
\end{equation*}
The filtration defined by \eqref{columnwise sigma algebra} moves column-wise through all the rows starting with column $j=0$ and ending with column $j=J-1$ (outputting column $J$), that is, we complete the yellow column $j+1$ in Figure \ref{Figure2}, before moving to the next column by using the identity ${\cal H}_{I,j}={\cal B}_{j+1}= {\cal H}_{0,j+1}$.

For the latent state-space processes $\Theta_{1:I,j}$, we consider the analogous flow of information
\begin{equation*}
  {\cal G}_{i,j} = \sigma \left( \Theta_{1:I,0:j-1}\right) \vee \sigma \left( \Theta_{1:i,j}\right).
\end{equation*}
The $\sigma$-fields ${\cal H}_{i,j}$ consider the observable cumulative claims, the $\sigma$-fields ${\cal G}_{i,j}$ are auxiliary latent objects (unobservable) that are needed to formulate our observation-driven state-space model. This model is going to be based on gamma distributed random variables $X\sim \Gamma(\gamma, c)$ with shape parameter $\gamma>0$ and rate parameter $c>0$ resulting in mean $\E[X]=\gamma/c$. Moreover, we mostly consider reciprocal gamma random variables with mean $\E[X^{-1}]=c/(\gamma-1)$ for $\gamma>1$.

\subsection{Observation-driven gamma-gamma state-space CL model}

We introduce a state-space model for every development period $j+1 \in \{1,\ldots, J\}$. This state-space model is based on a minimal set of assumptions. There will be an extended discussion after these assumptions. 


\begin{ass}[Observation-driven gamma-gamma state-space CL model]\label{model assumptions}
Fix the following set of deterministic parameters: $\gamma_j>0$, $f_j>0$, $\sigma_j^2>0$, $p_{i,j}\in [0,1]$ and $q_{i,j} >0$ for $i\in \{1,\ldots, I-1\}$ and $j\in \{0,\ldots, J-1\}$.  

\bigskip

{\underline{\bf (1) State-space equation.}} For $j \in \{0,\ldots, J-1\}$ and $i \in \{1,\ldots, I\}$, assume 
\begin{equation}\label{state-space}
\left. \Theta_{i,j} \right| {\cal H}_{i-1,j}~\sim~ 
\Gamma \left(1+a_{i,j}^-, \,
b_{i,j}^-\right),
\end{equation}
with initialization $a_{1,j}^-=\gamma_j>0$ and $b_{1,j}^-=f_j\gamma_j>0$, and affine updates for $i\in \{1,\ldots, I-1\}$
\begin{eqnarray}\label{recursion formula 1}
a_{i+1,j}^-&=&\left(p_{i,j}+q_{i,j}\right)a_{i,j}^+,\\
  b_{i+1,j}^-&=& p_{i,j}\,a_{i,j}^+\,f_j+ q_{i,j}\,b_{i,j}^+,\label{recursion formula 2}
\end{eqnarray}                  
where we defined for $i\in \{1,\ldots, I\}$
\begin{equation}\label{recursion formula a plus}
a_{i,j}^+=a_{i,j}^-+{C_{i,j}}/{\sigma_j^2}~>~0 \qquad \text{ and } \qquad
  b_{i,j}^+= b_{i,j}^-+{C_{i,j+1}}/{\sigma_j^2}~>~0.
\end{equation}

\bigskip

{\underline{\bf (2) Observation equation.}} For $i \in \{1,\ldots, I\}$ and $j\in \{0,\ldots, J-1\}$, we assume for the cumulative claims
\begin{equation}\label{observation equation}
\left. C_{i,j+1} \right|{\cal H}_{i-1,j}, {\cal G}_{i,j}~\stackrel{\rm (d)}{=}~ 
\left. C_{i,j+1} \right|C_{i,j}, \Theta_{i,j}~\sim~
\Gamma \left(\frac{C_{i,j}}{\sigma_j^2}, \,
\frac{\Theta_{i,j}}{\sigma_j^2}\right).
\end{equation}
Moreover, the initial column $C_{1:I,0}$ is always assumed to be fixed and strictly positive.
\end{ass}


\bigskip

We discuss Assumptions \ref{model assumptions}:
\begin{itemize}
\item The conditional law of the cumulative claim $C_{i,j+1}$ in \eqref{observation equation} is fully determined by the latent state $\Theta_{i,j}$ and the previous cumulative claim $C_{i,j}$. It has conditional mean
  \begin{equation}\label{CL property 1}
    \E \left[ C_{i,j+1} \middle| {\cal H}_{i-1,j}, {\cal G}_{i,j} \right]=
    \E \left[ C_{i,j+1} \middle| C_{i,j}, \Theta_{i,j}\right] = \Theta^{-1}_{i,j}\, C_{i,j}.
  \end{equation}
  Thus, the reciprocal latent state $\Theta^{-1}_{i,j}$ plays the role of an {\it individual CL factor}.
  Moreover, we obtain Mack's \cite{Mack} linear conditional variance assumption for given individual CL factors
  \begin{equation}\label{CL property 1PLUS}
    \operatorname{Var} \left( C_{i,j+1} \middle| {\cal H}_{i-1,j}, {\cal G}_{i,j} \right)=
    \operatorname{Var} \left( C_{i,j+1} \middle| C_{i,j}, \Theta_{i,j}\right) = \sigma_j^2\Theta^{-2}_{i,j}\, C_{i,j},
  \end{equation}
  with variance parameter $\sigma_j^2\Theta^{-2}_{i,j}>0$.
\item In Lemma \ref{cor mean-stationarity}, below, we see that $\Theta^{-1}_{1:I,j}$ is mean-stationary in accident year $i$ with $\E[\Theta^{-1}_{i,j}]=f_j$ for all $i\in \{1,\ldots, I\}$. Thus, each individual CL factor in \eqref{CL property 1} has {\it baseline mean} $f_j>0$. We are going to study below how this baseline assessment changes through Bayesian inference.
\item As highlighted in Ahn et al.~\cite[Remark 2]{Ahn2}, there exist infinitely many stochastic models that comply with Assumptions \ref{model assumptions}, e.g., if one uses the gamma-thinning construction presented in Ahn et al.~\cite{Ahn1}, one receives a full stochastic model that has the affine parameter structure \eqref{recursion formula 1}-\eqref{recursion formula a plus} under the additional assumption of $p_{i,j}q_{i,j}\in (0,1)$. We emphasize that Assumptions \ref{model assumptions} do not uniquely define a model, i.e., the state-space update \eqref{state-space} does not uniquely define the state-space dynamics, but rather a large model class thereof. In particular, it allows for different covariance structures in the reciprocal latent state-space processes $\Theta^{-1}_{1:I,j}$; see Ahn et al.~\cite{Ahn2, Ahn3}. This is further discussed below.
\end{itemize}

\section{Filtering, credibility, and mean-stationarity}
\label{Filtering, credibility, and mean-stationarity}
We discuss the different types of CL factors that result from Assumptions \ref{model assumptions}. For this,
we introduce the following notation
\begin{equation*}
f_{i,j}^-=\frac{b_{i,j}^-}{a_{i,j}^-}, \quad  
f_{i,j}^+=\frac{b_{i,j}^+}{a_{i,j}^+}, \quad
\omega_{i,j}=\frac{C_{i,j}/\sigma_j^2}{a_{i,j}^-+C_{i,j}/\sigma_j^2}\in(0,1), \quad \text{and} \quad
\Delta_{i,j}= \frac{ q_{i,j}}{p_{i,j}+q_{i,j}}\in(0,1].
\end{equation*}
\subsection{Bayesian update and a first credibility formula}
Based on the observation equation \eqref{observation equation}, having observed $C_{i,j+1}$, we apply a Bayesian inference step. Note that ${\cal H}_{i,j}={\cal H}_{i-1,j}\vee \sigma(C_{i,j+1})$, and applying Bayes' rule to
\eqref{observation equation} yields
\begin{equation}\label{Baysian inference i1}
\left. \Theta_{i,j} \right| {\cal H}_{i,j}~\sim~ 
\Gamma \left(1+a_{i,j}^+,~b_{i,j}^+\right).
\end{equation}
The state-space equation \eqref{state-space} gives us the {\it predictive prior mean}, based on history ${\cal H}_{i-1,j}$,
\begin{equation}\label{prior mean 0}
f^-_{i,j}~=~ \E \left[ \Theta_{i,j}^{-1} \middle| {\cal H}_{i-1,j}\right]~=~ \frac{b_{i,j}^-}{a_{i,j}^-},
\end{equation}
and the Bayesian inference step \eqref{Baysian inference i1} turns this into the {\it posterior mean}
\begin{equation}\label{credibility 1}
f^+_{i,j}~=~ \E \left[ \Theta_{i,j}^{-1} \middle| {\cal H}_{i,j}\right]
  ~=~ \frac{b_{i,j}^+}{a_{i,j}^+} ~=~ \omega_{i,j} \,\frac{C_{i,j+1}}{C_{i,j}} + \left(1-\omega_{i,j}\right)f^-_{i,j}.
\end{equation}
Thus, the Bayesian inference step yields the B\"uhlmann--Straub \cite{BS} credibility form for the individual CL factor $\Theta_{i,j}^{-1}$, conditional on the observations ${\cal H}_{i,j}$, with predictive prior mean $f^-_{i,j}$, observation update ${C_{i,j+1}}/{C_{i,j}}$
and credibility weight $\omega_{i,j}\in (0,1)$.

\subsection{State-space update and a second credibility formula}
\label{Bayesian filtering}
Based on \eqref{Baysian inference i1}, we consider the state-space update \eqref{state-space} for $i \to i+1$. That is,
\begin{equation}\label{state space update explicit}
\left. \Theta_{i,j} \right| {\cal H}_{i,j}~\sim~ 
\Gamma \left(1+a_{i,j}^+,~b_{i,j}^+\right)
\qquad \longrightarrow \qquad
\left. \Theta_{i+1,j} \right| {\cal H}_{i,j}~\sim~ 
\Gamma \left(1+a_{i+1,j}^-, \,
b_{i+1,j}^-\right).
\end{equation}
This is precisely the step that lacks uniqueness in the model definition of Assumptions \ref{model assumptions}.
Both conditional distributions of $\Theta_{i,j}$ and $\Theta_{i+1,j}$ are conditioned on the same information set ${\cal H}_{i,j}$, i.e., we consider the conditional joint distribution of
\begin{equation}\label{missing copula}
\left. \Theta_{i,j}, \Theta_{i+1,j} \right| {\cal H}_{i,j},
\end{equation}
in the state-space update \eqref{state space update explicit} that maps between the two conditional gamma distributions. Obviously, any copula can be selected in \eqref{missing copula} to link these two conditional distributions. Therefore, we face the problem of non-uniqueness. For the computation of ultimate claim predictions and the conditional MSEP however, this non-identifiability is irrelevant (this is seen below), it only starts to matter if we try to understand the risk drivers in the conditional MSEP.

We compute the conditional mean of the state-space update, i.e., the new predictive prior mean, see \eqref{prior mean 0}. This uses the assumptions \eqref{recursion formula 1}-\eqref{recursion formula a plus} to obtain
\begin{eqnarray}\nonumber
f^-_{i+1,j}&=&
  \E \left[ \Theta^{-1}_{i+1,j}\middle| {\cal H}_{i,j} \right] 
~=~ \frac{b_{i+1,j}^-}{a_{i+1,j}^-} 
  \\&=&\label{credibility 2}
   \Delta_{i,j}\, f_{i,j}^+
        +\left(1-\Delta_{i,j}\right) f_j
        \\&=&\nonumber
\Delta_{i,j}\left(\omega_{i,j} \,\frac{C_{i,j+1}}{C_{i,j}} + \left(1-\omega_{i,j}\right)f^-_{i,j}\right)
+\left(1-\Delta_{i,j}\right) f_j.
\end{eqnarray}
The nature of the first credibility formula \eqref{credibility 1} is different from the second one
\eqref{credibility 2}. The former is a proper B\"uhlmann credibility form from a Bayesian inference step, whereas the latter is an enforced one by the affine parameter assumptions \eqref{recursion formula 1}-\eqref{recursion formula a plus} in the state-space update \eqref{state-space}.

\subsection{Mean-stationarity}
A main purpose of setting up the affine mean structure in the state-space update \eqref{state-space} is to receive the following mean-stationarity of the latent process. In particular, a priori, before having any observations, we cannot differentiate the expected CL factors between the different accident years $i$. The proof of the following lemma is given in the appendix.

\begin{lemma}\label{cor mean-stationarity}
  The above model satisfies mean-stationarity $\E [ \Theta^{-1}_{i,j}| {\cal B}_j]=f_j$
  for all $i\in \{1,\ldots, I\}$.
\end{lemma}

This mean-stationarity is based on the following observations  used in the proof
of Lemma \ref{cor mean-stationarity} in the appendix:
\begin{eqnarray}\label{used later on}
  \E\left[C_{i,j+1} \middle| {\cal H}_{i-1,j}\right] 
&=&  \E\left[  \E\left[C_{i,j+1} \middle| {\cal H}_{i-1,j}, {\cal G}_{i,j}\right] 
    \middle| {\cal H}_{i-1,j}\right]
\\&=& C_{i,j} \, \E\left[  \Theta^{-1}_{i,j} \middle| {\cal H}_{i-1,j}\right]     
      ~=C_{i,j} \, f_{i,j}^-,
      \nonumber
\end{eqnarray}
the right-hand side considers the predictive prior mean $f_{i,j}^-$ of $\Theta^{-1}_{i,j}$, given ${\cal H}_{i-1,j}$, see \eqref{prior mean 0}.
Moreover, the tower property yields the calibration for the posterior mean, see \eqref{credibility 1},
\begin{equation}\label{posterior mean calibration}
  \E\left[f_{i,j}^+ \middle| {\cal H}_{i-1,j}\right]=f_{i,j}^-.
\end{equation}
This can be iterated which gives the following corollary.
\begin{cor} \label{corollary unbiasedness posterior}
  For $h\in \{0,\ldots, i-1\}$
  \begin{equation*}
    \E\left[f_{i,j}^- \middle| {\cal H}_{h,j}\right]=
    \E\left[f_{i,j}^+ \middle| {\cal H}_{h,j}\right]=
    \left(\prod_{k=h+1}^{i-1} \Delta_{k,j}\right) f_{h+1,j}^-
    + \left(1-\prod_{k=h+1}^{i-1} \Delta_{k,j}\right) f_j,
  \end{equation*}
where an empty product is set equal to one, and with initialization $f_{1,j}^-=f_j$.
\end{cor}

\begin{center}
\begin{tabularx}{\textwidth}{@{}lX@{}}
\toprule
\textbf{Symbol}&\textbf{Interpretation}\\\midrule
$\Theta_{i,j}^{-1}$& Individual CL factor.\\[3pt]
  $f_j$&Fixed baseline mean $\E[\Theta^{-1}_{i,j}]$ of the individual CL factor.\\[3pt]
$f^-_{i,j}$&Predictive prior mean $\E [ \Theta_{i,j}^{-1} | {\cal H}_{i-1,j}]$, before observing $C_{i,j+1}$.\\[3pt]
$f^+_{i,j}$&Posterior mean $\E [ \Theta_{i,j}^{-1} | {\cal H}_{i,j}]$, after observing $C_{i,j+1}$.\\[3pt]
$\widehat{f}^{(I)}_{i,j}$& Observable CL factor  $\E [ f_{i,j}^{-} | {\cal D}_{I}]$ at time $I$ for $i+j\ge I$, see \eqref{4 identities}.\\
\bottomrule
\end{tabularx}
\end{center}

\section{Ultimate claim predictions}
\label{Ultimate claim predictions}
We translate the above considerations to ultimate claim predictions of $C_{i,J}$, $i+J>I$, under trapezoidal observations ${\cal D}_I$
\begin{equation*}
  \widehat{C}_{i,J} = \E \left[C_{i,J} \middle| {\cal D}_I \right],
\end{equation*}
we refer to Figure \ref{Figure1} (lhs). Thus, we consider all accident years $i$ with incomplete row observations.

\subsection{Data likelihood}
\label{section data likelihood}
The available data ${\cal D}_I$ is either a triangle for $I=J+1$ or a trapezoid for $I>J+1$. The data likelihood factorizes along the development year axis $j$, and it can be written in an elegant way along the accident year axis $i$.

We start by introducing the full data distribution over the entire development rectangle. We therefore introduce the (scaled) prime-beta densities on the positive support $x>0$
  \begin{equation*}
    K\left(x\middle| c, a, b\right)
    = \frac{\Gamma(c+1+a)}{\Gamma(c)\Gamma(1+a)}
    \frac{b^{1+a}\, x^{c-1} }{(b+x)^{c+1+a}},
   \end{equation*} 
   for positive parameters $a,b,c>0$.
   The prime-beta random variables $C\sim K(\,\cdot\mid c,a,b)$ can be simulated from independent random variables $U\sim\Gamma(c,1)$ and $V\sim\Gamma(1+a,b)$ by setting $C=U/V$.

\begin{lemma}\label{likelihood lemma}
The conditional density of the model in Assumptions \ref{model assumptions}, given the first column $C_{1:I,0}$, factorizes as
  \begin{equation*}
    f\left((C_{i,j})_{i=1,\ldots, I,\, j=1,\ldots, J} \middle| C_{1:I,0}\right)
    = \prod_{j=0}^{J-1} \prod_{i=1}^{I}\,
        K\left(C_{i,j+1}\,\middle| \,C_{i,j}/\sigma_j^2, \,a_{i,j}^-,\, \sigma_j^2 b_{i,j}^-\right).
  \end{equation*}
\end{lemma}
The proof of this lemma is provided in the appendix.
We relate this lemma to the data likelihood of ${\cal D}_I$. There are two important observations. First, the trapezoidal structure respects parenting, meaning that $C_{i,j+1}$ only depends on
$C_{1:i,j}$ and $C_{1:i-1,j+1}$. Second, related to the first item, the prime-beta densities only depend on these parents, but not on the entire history ${\cal H}_{i-1,j}$, that is, 
\begin{equation}\label{relevant history}
K\left(C_{i,j+1}\,\middle| \,C_{i,j}/\sigma_j^2, \,a_{i,j}^-,\, \sigma_j^2 b_{i,j}^-\right)
=f\left(C_{i,j+1}\,\middle| \,C_{1:i,j}, C_{1:i-1,j+1}\right).
\end{equation}
These two items allow us to adapt the density of Lemma \ref{likelihood lemma} to the trapezoidal shape of data ${\cal D}_I$ by first integrating out
$C_{r_J+1:I,J}$, then $C_{r_{J-1}+1:I,J-1}$, etc. This results in Corollary \ref{data likelihood lemma}, below, but we first introduce some notation.

\begin{figure}[htb!]
\begin{center}
\begin{minipage}[t]{0.5\textwidth}
\begin{center}
\includegraphics[width=\textwidth]{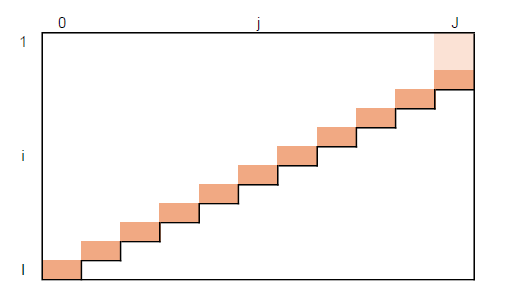}
\end{center}
\end{minipage}
\end{center}
\vspace{-.7cm}
\caption{Latest observations $C_{i,d_i}$ and $C_{r_j, j}$ at time $I$ for $i \in \{1,\ldots, I\}$ and $j \in \{0,\ldots, J\}$.}
\label{Figure2B}
\end{figure}

The index $d_i \in \{0,\ldots, J\}$ in \eqref{index di} defines the latest observed cumulative claims $C_{i,d_i}$ at time $I$ for all accident years $i\in \{1,\ldots, I\}$. Equivalently, we can define the latest observations $C_{r_j,j}$ for all columns 
$j\in \{0,\ldots, J\}$ by the accident year indices
\begin{equation*}
r_j=I-j \in \{I-J,\ldots, I\}.
\end{equation*}
These two indices are illustrated in Figure \ref{Figure2B}. This provides the following corollary from Lemma \ref{likelihood lemma}, for convenience, we shift the development parameter $j$ by one unit.

\begin{cor}\label{data likelihood lemma}
  The conditional data likelihood of ${\cal D}_I$, given the first column $C_{1:I,0}$, factorizes as
  \begin{equation*}
    f\left((C_{i,j})_{i+j \le I,\, j\ge 1} \middle| C_{1:I,0}\right)
    = \prod_{j=1}^{J} \prod_{i=1}^{r_j}
        K \left(C_{i,j}\middle| \, C_{i,j-1}/\sigma^2_{j-1}, \, a_{i,j-1}^-, \, \sigma^2_{j-1}b_{i,j-1}^-\right).
   \end{equation*}
\end{cor}
The likelihoods of Lemma \ref{likelihood lemma} and Corollary \ref{data likelihood lemma} give nice insights. We are able to express the data likelihood explicitly (fully analytically), and the non-identifiability of the copula in \eqref{missing copula} does not have any impact. Thus, Assumptions \ref{model assumptions} are sufficient to compute the ultimate claim predictions and the prediction uncertainty, given the data ${\cal D}_I$.
We can interpret the latent state-space processes as auxiliary tools that help us to shape our understanding and interpretation, and to arrive at the likelihoods of Lemma \ref{likelihood lemma} and Corollary \ref{data likelihood lemma}.

Another noteworthy remark is that the accident year factorization
in Lemma \ref{likelihood lemma} and Corollary \ref{data likelihood lemma} is not an independent one as $a_{i,j}^-$ and $b_{i,j}^-$ collect the observed histories $C_{1:i-1,j}$ and $C_{1:i-1,j+1}$.

\subsection{Observable chain-ladder factors and ultimate claim prediction}
\begin{figure}[htb!]
\begin{center}
\begin{minipage}[t]{0.95\textwidth}
\begin{center}
\includegraphics[width=\textwidth]{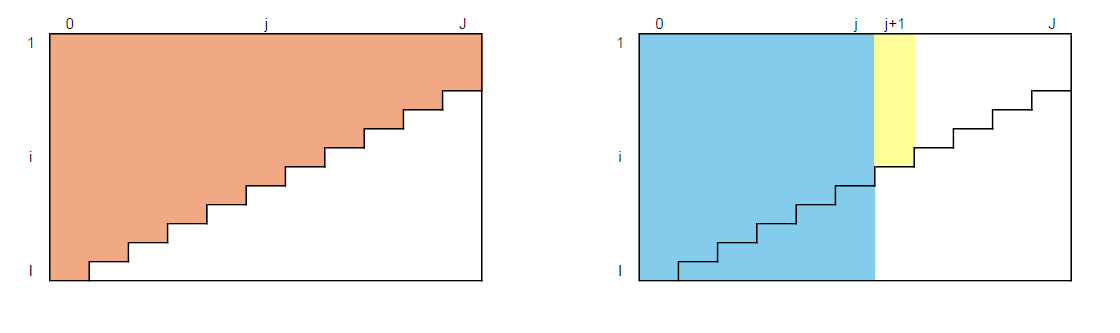}
\end{center}
\end{minipage}
\end{center}
\vspace{-.7cm}
\caption{(lhs) Information set ${\cal D}_I$ and (rhs) information set
${\cal H}_{r_j-1,j}={\cal H}_{r_{j+1},j}$.}
\label{Figure3}
\end{figure}

To compute the ultimate claim predictions, we need to be able to determine the CL factors, given ${\cal D}_I$. Such a computation has already been presented in Corollary \ref{corollary unbiasedness posterior}, however, the
information sets ${\cal D}_I$ and ${\cal H}_{r_j-1,j}$ do not directly align, see Figure \ref{Figure3}.

We compute for a fixed $j \in \{0,\ldots, J-1\}$
\begin{equation*}
\E \left[ f_{i,j}^{-} \middle| {\cal D}_I \right]
\qquad \text{ for $i = r_j, \ldots, I$},
\end{equation*}
this corresponds to the cumulative claims $C_{i,j+1}$, see \eqref{used later on}, which are illustrated by the continuation of the yellow column in Figure \ref{Figure3} (rhs). Fortunately, this can be done, and the result is provided in the next proposition; the proof is given in the appendix.

\begin{prop} \label{Chain Ladder Factor Proposition}
Consider $i\ge r_j= I-j$. We have observable CL factors at time $I$
  \begin{equation*}
\widehat{f}^{(I)}_{i,j} = \E\left[f_{i,j}^- \middle| {\cal D}_{I}\right]=   \E\left[f_{i,j}^- \middle| {\cal H}_{r_j-1,j}\right]=
    \left(\prod_{k=r_j}^{i-1} \Delta_{k,j}\right) f_{r_j,j}^-
    + \left(1-\prod_{k=r_j}^{i-1} \Delta_{k,j}\right) f_j.
  \end{equation*}
\end{prop}

Using \eqref{credibility 2} and the conditional independent decoupling argument of the previous proof, we also get
  \begin{equation}\label{4 identities}
\widehat{f}^{(I)}_{i,j} = \E\left[f_{i,j}^- \middle| {\cal D}_{I}\right]=   \E\left[f_{i,j}^- \middle| {\cal H}_{r_j-1,j}\right]=   \E\left[\Theta_{i,j}^{-1} \middle| {\cal H}_{r_j-1,j}\right].
  \end{equation}
  Thus, $\widehat{f}^{(I)}_{i,j}$ is the observable CL factor at time $I$.
Proposition \ref{Chain Ladder Factor Proposition} now allows us to compute the ultimate claim predictions, given observations ${\cal D}_I$.
We have the following theorem.

\begin{theo}\label{CL forecast}
For accident years $i+J > I$ we have ultimate claim predictor
\begin{equation*}
  \widehat{C}_{i,J}
  ~=~ \E \left[C_{i,J} \middle| {\cal D}_I \right]
~=~C_{i,d_i}\, \prod_{j=d_i}^{J-1}
\widehat{f}^{(I)}_{i,j}.
\end{equation*}
\end{theo}
The proof of this theorem is given in the appendix.
For accident years $i+J> I$, Theorem \ref{CL forecast} requires computation of the observable CL factors, see Proposition \ref{Chain Ladder Factor Proposition},
\begin{equation}\label{observable CL factor update}
\widehat{f}^{(I)}_{i,j} =
    \left(\prod_{k=r_j}^{i-1} \Delta_{k,j}\right) f_{r_j,j}^-
    + \left(1-\prod_{k=r_j}^{i-1} \Delta_{k,j}\right) f_j.
\end{equation}
These observable CL factors involve the deterministic weights
$\Delta_{k,j}=q_{k,j}/(p_{k,j}+q_{k,j}) \in (0,1]$, the fixed baseline mean $f_j$, 
and the observable predictive prior mean at time $I$ given by
\begin{equation}\label{full update formula}
f_{r_j,j}^- =
\Delta_{r_j-1,j}\left(\omega_{r_j-1,j} \,\frac{C_{r_j-1,j+1}}{C_{r_j-1,j}} + \left(1-\omega_{r_j-1,j}\right)f^-_{r_j-1,j}\right)
+\left(1-\Delta_{r_j-1,j}\right) f_j.
\end{equation}
with weights
\begin{eqnarray*}
  \omega_{r_j-1,j}&=&\frac{C_{r_j-1,j}/\sigma_j^2}{a_{r_j-1,j}^-+C_{r_j-1,j}/\sigma_j^2}\\
  &=&\frac{C_{r_j-1,j}}{\sum_{k=1}^{r_j-1} C_{k,j}\prod_{\ell=k}^{r_j-2}\left(p_{\ell,j}+q_{\ell,j}\right) 
    + \sigma_j^2\gamma_j \prod_{\ell=1}^{r_j-2}\left(p_{\ell,j}+q_{\ell,j}\right) }~\in~(0,1).
\end{eqnarray*}
The latter uses \eqref{recursion formula 1} and \eqref{recursion formula a plus} to unfold $a_{i,j}^->0$
as follows
\begin{eqnarray}\nonumber
  a_{i,j}^- &=&\left(p_{i-1,j}+q_{i-1,j}\right){C_{i-1,j}}/{\sigma_j^2}
           +\left(p_{i-1,j}+q_{i-1,j}\right)a_{i-1,j}^-
  \\   &=&\sum_{k=1}^{i-1} \frac{C_{k,j}}{\sigma_j^2}\prod_{\ell=k}^{i-1}\left(p_{\ell,j}+q_{\ell,j}\right) 
           +\gamma_j \prod_{\ell=1}^{i-1}\left(p_{\ell,j}+q_{\ell,j}\right) .
 \label{unfold a minus}          
\end{eqnarray}
Formula \eqref{full update formula} gives the full updating formula of the observable CL factor across the accident year indices $i=2,\ldots, r_j$, this is the part that is covered by observations ${\cal D}_I$. Formula
\eqref{observable CL factor update} extrapolates the accident year index updates $i=r_j+1,\ldots, I$ into the lower triangle, where no observations are available. The structure is the same, but the missing observations are integrated out.

\subsection{Special case: Exponential seniority weighting in the CL factors}
\label{Special case section}
In the applications, our main interest lies in the special case $p_{i,j} \equiv 0$ and $q_{i,j} \equiv q_j\le 1$
The first assumption gives us $\Delta_{i,j}\equiv 1$ which
simplifies \eqref{full update formula} as follows, note that these observable CL factors no longer have an accident year $i$ dependence,
\begin{eqnarray}\nonumber
\widehat{f}^{(I)}_{j} &:=&
    f_{r_j,j}^- ~=~
\omega_{r_j-1,j} \,\frac{C_{r_j-1,j+1}}{C_{r_j-1,j}} + \left(1-\omega_{r_j-1,j}\right)f^-_{r_j-1,j}
\\&=& \label{observable CL factor update uniformative}
\sum_{k=1}^{r_j-1} \left[\prod_{\ell=k+1}^{r_j-1}
\left(1-\omega_{\ell,j}\right)\right]
\omega_{k,j} \,\frac{C_{k,j+1}}{C_{k,j}}
+ \prod_{\ell=1}^{r_j-1}
\left(1-\omega_{\ell,j}\right) f_j.
\end{eqnarray}
This is a classical credibility formula that stems from a time-series consideration, and the credibility weights
simplify under $q_{i,j} \equiv q_j\le 1$ to
\begin{equation}\label{comming back later to}
  \omega_{i,j}=\frac{C_{i,j}}{\sum_{k=1}^{i} q_j^{i-k} C_{k,j} 
    + q_j^{i-1} \sigma_j^2\gamma_j}~\in~(0,1).
\end{equation}
If we furthermore consider the non-informative prior limits $\gamma_j \downarrow 0$, we obtain
\begin{equation}\label{non-informative prior limit}
  \omega_{i,j}=\lim_{\gamma_j \downarrow 0}\frac{C_{i,j}}{\sum_{k=1}^{i} q_j^{i-k} C_{k,j} 
    + q_j^{i-1} \sigma_j^2\gamma_j}
  =\frac{C_{i,j}}{\sum_{k=1}^{i} q_j^{i-k} C_{k,j}}~\in~(0,1].
\end{equation}
Compared to the classical CL factors $\widehat{f}^{\rm CL}_j$ of Mack \cite{Mack}, we obtain an exponential seniority weighting $q_j^{i-k}$ in past observations; the classical CL factors of Mack \cite{Mack} are given by
\begin{equation}\label{classical Mack CL factors}
  \widehat{f}^{\rm CL}_j =\frac{\sum_{k=1}^{r_j-1} C_{k,j+1}}{\sum_{k=1}^{r_j-1} C_{k,j} }
=  \sum_{i=1}^{r_j-1} \left[\frac{ C_{i,j}}{\sum_{k=1}^{r_j-1} C_{k,j} }\right] \frac{C_{i,j+1}}{C_{i,j}}.
\end{equation}
Moreover, in this non-informative prior limit \eqref{non-informative prior limit},
also the credibility weight on the baseline $f_j$ in \eqref{observable CL factor update uniformative}
will vanish. In summary, this gives us the following theorem.

\begin{theo}[Exponential seniority weighted CL factors]\label{main theorem}
Under Assumptions \ref{model assumptions}, let $p_{i,j} \equiv 0$, $q_{i,j} \equiv q_j\le 1$ and $\gamma_j \to 0$. For accident years $i\ge I-J+1=r_{J-1}$, we have the ultimate claims predictors
\begin{equation*}
  \widehat{C}_{i,J}^{{\rm CL}(q)}
  ~:=~ \lim_{\bgamma \downarrow \b0}\, \E_{\bgamma} \left[C_{i,J} \middle| {\cal D}_I \right]
~=~C_{i,d_i}\, \prod_{j=d_i}^{J-1}
\widehat{f}^{(I)}_{j},
\end{equation*}
with CL factors and credibility weights, respectively, for $j \in \{0,\ldots, J-1\}$
\begin{equation}\label{CL factor for examples}
  \widehat{f}^{(I)}_{j} = \sum_{i=1}^{r_j-1} \alpha_{i,j} \,\frac{C_{i,j+1}}{C_{i,j}}
  \qquad \text{ and } \qquad
  \alpha_{i,j}=
  \frac{q_j^{r_j-1-i} C_{i,j}}{\sum_{k=1}^{r_j-1} q_j^{r_j-1-k} C_{k,j}}~\in~(0,1].
\end{equation}
\end{theo}

\begin{figure}[htb!]
\begin{center}
\begin{minipage}[t]{0.95\textwidth}
\begin{center}
\includegraphics[width=\textwidth]{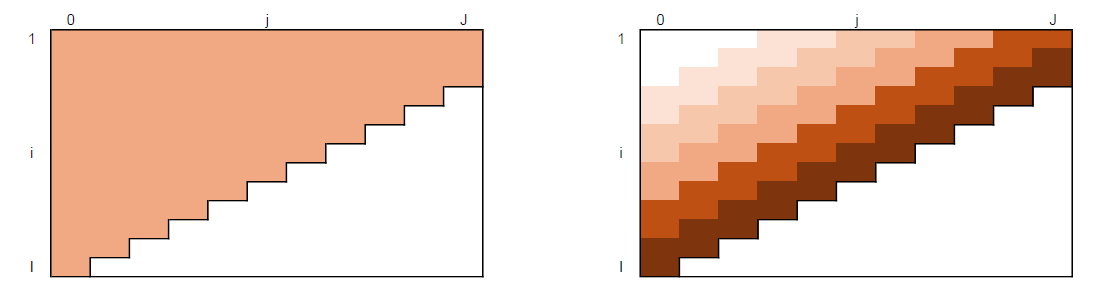}
\end{center}
\end{minipage}
\end{center}
\vspace{-.7cm}
\caption{(lhs) Information set ${\cal D}_I$ and (rhs) exponential seniority weighting
of past information according to Theorem \ref{main theorem}.}
\label{Figure4}
\end{figure}

Theorem \ref{main theorem} allows for an exponential seniority weighting in the computation of the CL factors, this is illustrated in Figure \ref{Figure4} (rhs). Moreover, the choice $q_j\equiv 1$, yields the classical CL predictors of Mack \cite{Mack}.

\section{Total prediction uncertainty}
\label{Total prediction uncertainty}
We introduce the following abbreviations for the accident year aggregated versions
\begin{equation*}
    C_\bullet=\sum_{i=1}^{I}C_{i,J},
    \qquad \text{ and } \qquad
    \widehat{C}_\bullet =\sum_{i=1}^{I}\widehat C_{i,J}
    =\E\left[C_\bullet\middle| {\cal D}_I\right].
\end{equation*}
We measure prediction uncertainty in the classical actuarial view by computing the conditional MSEP given by
\begin{equation*}
  \operatorname{msep}_{C_\bullet|{\cal D}_I}\left(\widehat{C}_\bullet\right)
  = \E \left[ \left(C_\bullet-\widehat{C}_\bullet\right)^2 \middle| {\cal D}_I\right].
\end{equation*}
Because we use the conditional expectations $\widehat{C}_{i,J}=\E[C_{i,J}|{\cal D}_I]$ as predictors,
we obtain
\begin{equation}\label{conditional MSEP}
\operatorname{msep}_{C_\bullet|{\cal D}_I}\left(\widehat{C}_\bullet\right)
  = \operatorname{Var}\left( C_\bullet \middle| {\cal D}_I\right).
\end{equation}
Coming back to the non-identifiability of the unspecified copula in \eqref{missing copula}, we have seen in
Section \ref{section data likelihood} that the unspecified copula does not impact the computation of the conditional MSEP. However, it can affect the analysis of the two terms of average process uncertainty and parameter uncertainty, respectively,
\begin{equation*}
\operatorname{msep}_{C_\bullet|{\cal D}_I}\left(\widehat{C}_\bullet\right)
= \E\left[\operatorname{Var}\left(C_\bullet \middle| \bTheta, {\cal D}_I\right)\middle| {\cal D}_I\right]+\operatorname{Var}\left( \E\left[ C_\bullet \middle| \bTheta, {\cal D}_I\right]
  \middle| {\cal D}_I\right).
\end{equation*}
here $\bTheta= \Theta_{1:I, 0:J-1}$. Our focus is on \eqref{conditional MSEP}, which means that we do not need to specify the copula.

\medskip

Unfortunately, we cannot compute \eqref{conditional MSEP} in closed-form, except for special cases like in the gamma-gamma Bayes' CL model of W\"uthrich--Merz \cite{WM2015}. Fortunately, the fact that we have a full likelihood from Lemma \ref{likelihood lemma} permits a Monte Carlo (MC) solution because we can simulate the lower triangle, given ${\cal D}_I$. That is, we can recursively simulate from
\begin{equation*}
    f\left((C_{i,j})_{i+j>I} \middle| {\cal D}_I\right)
    = \prod_{j=1}^{J} \prod_{i=r_j+1}^{I}\,
        K\left(C_{i,j}\,\middle| \,C_{i,j-1}/\sigma_{j-1}^2, \,a_{i,j-1}^-,\, \sigma_{j-1}^2 b_{i,j-1}^-\right),
  \end{equation*}
  and using the recursive structure \eqref{recursion formula 1}-\eqref{recursion formula a plus}
  with initialization $a_{1,j}^-=\gamma_j>0$ and $b_{1,j}^-=f_j\gamma_j>0$, and recall $r_j=I-j$.

\begin{algorithm}[htb!]
\caption{Conditional MC simulation of the lower triangle and MSEP estimation}
\label{algorithm conditional msep}
\begin{algorithmic}[1]
  \Require Observations $\{C_{i,j}:i+j\leq I\}$; for $j \in \{1,\ldots, J\}$, the parameters of Assumptions \ref{model assumptions}, $a_{r_j+1,j-1}^-$ and $b_{r_j+1,j-1}^-$; number of simulations $M>1$; predictor $\widehat{C}_\bullet$.
\Ensure MC estimate $\widehat{\operatorname{MSEP}}_M$ and MC error $\widehat{C}_\bullet-\overline{C}_\bullet$.

\For{$m=1,\ldots,M$}
    \State Set $C_{i,d_i}^{(m)}\gets C_{i,d_i}$ for every observed cell $i \in \{1,\ldots, I\}$.
    \For{$j=1,\ldots,J$}
        \For{$i=r_j+1,\ldots,I$}
                \State Draw
                $\displaystyle
                C_{i,j}^{(m)}\sim
                K\left(\,\cdot\,\middle| {C_{i,j-1}^{(m)}}/{\sigma_{j-1}^{2}},\,a^-_{i,j-1},\,\sigma_{j-1}^{2}b^-_{i,j-1}
                \right)$.
            \If{$i<I$}
                \State $\displaystyle
                    a^+_{i,j-1}\gets
                    a^-_{i,j-1}
                    +{C_{i,j-1}^{(m)}}/{\sigma_{j-1}^{2}}$.
                \State $\displaystyle
                    b^+_{i,j-1}\gets
                    b^-_{i,j-1}
                    +{C_{i,j}^{(m)}}/{\sigma_{j-1}^{2}}$.
                \State $\displaystyle
                    a^-_{i+1,j-1}\gets
                    (p_{i,j-1}+q_{i,j-1})a^+_{i,j-1}$.
                \State $\displaystyle
                    b^-_{i+1,j-1}\gets
                    p_{i,j-1}a^+_{i,j-1}f_{j-1}
                    +q_{i,j-1}b^+_{i,j-1}$.
            \EndIf
        \EndFor
    \EndFor
    \State Compute $\displaystyle C_\bullet^{(m)}\gets\sum_{i=1}^{I}C_{i,J}^{(m)}$.
    \EndFor
\State Compute $\displaystyle
    \overline{C}_\bullet
    \gets\frac{1}{M}\sum_{m=1}^{M}
                C_\bullet^{(m)}$.    
\State Compute $\displaystyle
    \widehat{\operatorname{MSEP}}_M
    \gets\frac{1}{M-1}\sum_{m=1}^{M}
                (C_\bullet^{(m)}-\overline{C}_\bullet)^2$.
\State \Return  $\widehat{\operatorname{MSEP}}_M$ and $\widehat{C}_\bullet-\overline{C}_\bullet$.
\end{algorithmic}
\end{algorithm}

The resulting algorithm is given in Algorithm \ref{algorithm conditional msep}.
If $\E[C_\bullet^2|{\cal D}_I]<\infty$, the MC simulation of Algorithm \ref{algorithm conditional msep} converges to $\operatorname{Var}(C_\bullet|{\cal D}_I)$, a.s. Checking finite second moments is a bit tricky, we give a sufficient condition.
For $j\in\{0, \ldots, J-1\}$, consider $A_{r_j,j}=a^-_{r_j,j}$, and iterate
$A_{i+1,j}=(p_{i,j}+q_{i,j})A_{i,j}$ for $i \ge r_j$. Clearly $a^-_{i,j} \ge A_{i,j}$ for every simulated history in the lower triangle. We then require
\begin{equation*}
  \varepsilon := \min_{i+j \ge I} (A_{i,j}-1) >1.
\end{equation*}
This implies
\begin{eqnarray*}\nonumber
\E \left[\left(f^-_{i+1,j}\right)^2\middle| {\cal H}_{i-1,j}\right]&=&
                                                                      \left(\Delta_{i,j}f^-_{i,j}+\left(1-\Delta_{i,j}\right) f_j\right)^2 + \frac{\Delta_{i,j}^2\omega_{i,j}(f^-_{i,j})^2}{a_{i,j}^--1}
  \\&\le&  \left(1+\varepsilon^{-1}\right)(f^-_{i,j})^2 + f_j^2.
\end{eqnarray*}
This can be iterated and implies $\E[C_\bullet^2|{\cal D}_I]<\infty$.

\medskip

We emphasize that the parameters $\gamma_j$, $f_j$, $\sigma_j^2$, $p_{i,j}$ and $q_{i,j}$ are deterministic and fixed throughout.

\section{Claims development result}
\label{CDR prediction uncertainty}
This section studies the claims development result (CDR), which describes the one-year update of the ultimate claims view used in solvency considerations; see Merz--W\"uthrich \cite{MW2008}.
Introduce the available information at time $I+1$
\begin{equation*}
  {\cal D}_{I+1} = \sigma\left\{ C_{i,j};~i+j\le I+1, \, 1\le i \le I, \, 0 \le j \le J \right\}.
\end{equation*}
This is the run-off view, we retain the accident years $1,\ldots, I$, and add the observations on the calendar year diagonal $I+1$, i.e., no new accident year is introduced. The CDRs are then given by
\begin{equation*}
  \operatorname{CDR}_i = \E\left[C_{i,J}\middle| {\cal D}_{I}\right]
  -\E\left[C_{i,J}\middle| {\cal D}_{I+1}\right] \qquad \text{ and } \qquad  
  \operatorname{CDR} = \sum_{i=1}^I \operatorname{CDR}_i.
\end{equation*}
By the tower property, $\E[\operatorname{CDR}| {\cal D}_{I}]=0$.
We investigate the conditional MSEP of the CDR
\begin{equation}\label{simplifying CDR}
  \operatorname{msep}_{\operatorname{CDR}|{\cal D}_I}\left(0\right)
  = \E \left[ \operatorname{CDR}^2 \middle| {\cal D}_I\right]
      ={\rm Var}\left(\operatorname{CDR} \middle| {\cal D}_I\right)
=     {\rm Var}\left(\sum_{i=1}^I \E\left[C_{i,J}\middle| {\cal D}_{I+1}\right] \middle| {\cal D}_I\right). 
\end{equation}
Our goal is to compute this last expression.
\subsection{Prediction uncertainty in the one-year update}
We start by introducing the relevant notation. For $i\in \{I-J+1, \ldots, I\}$ and $j\in \{d_i+1, \ldots, J-1\}$, we abbreviate
\begin{equation*}
  \delta_{i,j}=\prod_{k=r_j}^{i-1}\Delta_{k,j} \qquad \text{ and } \qquad
  w_j=\omega_{r_j,j}=\frac{C_{r_j,j}/\sigma_j^2}{a_{r_j,j}^-+C_{r_j,j}/\sigma_j^2}
~\in~(0,1).
\end{equation*}
This allows us to define the following ${\cal D}_I$-measurable variables for $j>d_i$
\begin{equation*}
\nu_{i,j} = \delta_{i,j} (1-w_{j})f^-_{r_j,j} + (1-\delta_{i,j})f_j
  \qquad \text{ and } \qquad
  \kappa_{i,j}=\delta_{i,j} \frac{  w_{j}}{C_{r_j,j}}.
\end{equation*}
Moreover, we set the boundary definitions
\begin{equation*}
  (\nu_{i,j},\kappa_{i,j})
=\begin{cases}
  (0,1) & \text{ for $j=d_i$},\\
  (1,0) & \text{ for $j<d_i$}.
\end{cases}          
\end{equation*}
Under Assumptions \ref{model assumptions}, and in complete analogy to \eqref{observable CL factor update}-\eqref{full update formula}, the $(I+1)$-observable CL factors, for $i\ge I-J+1=r_{J-1}$  and $j>d_i$, are given by
\begin{equation*}
\widehat{f}^{(I+1)}_{i,j} =
    \left(\prod_{k=r_j+1}^{i-1} \Delta_{k,j}\right) f_{r_j+1,j}^-
                              + \left(1-\prod_{k=r_j+1}^{i-1} \Delta_{k,j}\right) f_j
=  \nu_{i,j} + \kappa_{i,j} \,C_{r_j,j+1}.   
\end{equation*}
This allows us to rewrite the ultimate claim predictors at time $I+1$ for $i\ge I-J+1$ by
\begin{equation}\label{new one-year representation}
\E \left[C_{i,J} \middle| {\cal D}_{I+1} \right]
~=~C_{i,d_i+1}\, \prod_{j=d_i+1}^{J-1}
\widehat{f}^{(I+1)}_{i,j}~=~ \prod_{j=0}^{J-1}\left(\nu_{i,j} + \kappa_{i,j} \,C_{r_j,j+1}\right).
\end{equation}
This one-year update $I\to I+1$ considers the diagonal observations
$(C_{r_j+1,j})_{j=1}^{J}$, and these are conditionally independent, given ${\cal D}_I$,
\begin{equation}\label{conditional independence}
    f\left((C_{r_j+1,j})_{j=1,\ldots, J} \middle| {\cal D}_I\right)
    = \prod_{j=1}^{J} 
        K\left(C_{r_j+1,j}\,\middle| \,C_{r_j+1,j-1}/\sigma_{j-1}^2, \,a_{r_j+1,j-1}^-,\, \sigma_{j-1}^2 b_{r_j+1,j-1}^-\right).
  \end{equation}
This follows directly from Lemma \ref{likelihood lemma}. This conditional independence property is the key for the closed-form solution of the conditional MSEP of the CDR.

Using similar arguments as in \eqref{used later on} yields
\begin{eqnarray}\label{eq 11}
\mu_j:=\E\left[C_{r_j,j+1} \middle| {\cal D}_{I}\right] 
&=& f_{r_j,j}^-\, C_{r_j,j},
 \\
\tau_j^2:= \E\left[C^2_{r_j,j+1} \middle| {\cal D}_{I}\right] 
&=& \left(f_{r_j,j}^-\right)^2\frac{a_{r_j,j}^-}{a_{r_j,j}^--1}\left(C^2_{r_j,j} + \sigma_j^2 C_{r_j,j}\right),
    \label{eq 22}
\end{eqnarray}
where the second line requires $a_{r_j,j}^->1$.

\begin{theo}\label{CDR theorem}
Under Assumptions \ref{model assumptions} and $a_{r_j,j}^->1$ for all $j\in \{0,\ldots, J-1\}$
\begin{equation*}
  \operatorname{msep}_{\operatorname{CDR}|{\cal D}_I}\left(0\right)
  = \sum_{i=I-J+1}^I\sum_{k=I-J+1}^I \left(\prod_{j=0}^{J-1} T_{i,k,j}-\widehat{C}_{i,J}\widehat{C}_{k,J}\right),
\end{equation*}
where the ultimate claims predictions are computed from Theorem \ref{CL forecast} and with
\begin{equation*}
T_{i,k,j}=  
\nu_{i,j}\nu_{k,j}+\left(\nu_{i,j}\kappa_{k,j}+\nu_{k,j}\kappa_{i,j}\right)\mu_{j} + \kappa_{i,j} \kappa_{k,j}\tau_j^2.
\end{equation*}  
\end{theo}
The key difference between the one-year MSEP and the total MSEP \eqref{conditional MSEP} is that in the former we obtain the conditional independence decoupling \eqref{conditional independence} which allows for closed-from solutions. This does independent decoupling, given ${\cal D}_I$, does not hold in the total MSEP case.

\subsection{The exponential seniority weighting case}
In the exponential seniority weighting case of Section \ref{Special case section}, we receive an interpretable version
of Theorem \ref{CDR theorem} that is in the spirit of W\"uthrich--Merz \cite[Theorem 2.15]{WM2015}.
That is, consider the special case $p_{i,j} \equiv 0$ and $q_{i,j} \equiv q_j \le 1$.
This implies, see \eqref{unfold a minus},
\begin{eqnarray*}\nonumber
  A_j~:=~a_{r_j,j}^- &=&\sum_{k=1}^{r_j-1} q_j^{r_j-k}\, \frac{C_{k,j}}{\sigma_j^2}
                       +  q_j^{r_j-1}\gamma_j,
  \\   B_j~:=~b_{r_j,j}^- &=&\sum_{k=1}^{r_j-1} q_j^{r_j-k}\, \frac{C_{k,j+1}}{\sigma_j^2}
                       +  q_j^{r_j-1}\gamma_j f_j,
\end{eqnarray*}
with $\widehat{f}_j^{(I)}=B_j/A_j$.
Introduce for $A_j>1$ the ${\cal D}_I$-measurable variables
\begin{eqnarray*}
\Psi_j&=& \frac{1}{A_j-1}
=\frac{\sigma_j^2}
{\sum_{k=1}^{r_j-1}q_j^{r_j-k}C_{k,j}+\sigma_j^2(q_j^{r_j-1}\gamma_{j}-1)},\\
  \omega^\star_j&=&\frac{C_{r_j,j}}{\sum_{\ell=1}^{r_j}q_j^{r_j-\ell}C_{\ell,j}+q_j^{r_j-1}\sigma_j^2 \gamma_j}
~=~ \frac{C_{r_j,j}}{\sigma_j^2 A_j + C_{r_j,j}},
\end{eqnarray*}
the latter stems from weight $\omega_{r_j,j}$ given in \eqref{comming back later to}.

\begin{theo} \label{theo WM CDR}
  Consider the special case $p_{i,j} \equiv 0$ and $q_{i,j} \equiv q_j \le 1$.
  For accident years $i+J >I$
\begin{equation*}
{\rm msep}_{\operatorname{CDR}_{i}|{\cal D}_I}(0)
=(\widehat{C}_{i,J})^2
\left[\left(1+\frac{\Psi_{d_i}}{\omega_{d_i}^\star}\right)
\prod_{j=d_i+1}^{J-1}
\left(1+
\omega_j^\star \Psi_{j}\right)-1\right],
\end{equation*}
where we assume that 
$A_j>1$ for all 
$ d_i \le j \le  J-1$, otherwise the corresponding conditional MSEP is infinite, and for 
aggregated accident years
\begin{equation*}
{\rm msep}_{\operatorname{CDR}|{\cal D}_I}(0)
  =\sum_i{\rm msep}_{\operatorname{\rm CDR}_{i}|{\cal D}_I}(0)
  +
   2\,\sum_{i<m}
\widehat{C}_{i,J}\,
\widehat{C}_{m,J}
\left[\left(1+\Psi_{d_i}\right)
 \prod_{j=d_i+1}^{J-1}
\left(1+\omega_j^\star \Psi_{j}\right)-1\right],
\end{equation*}
where the summations run over $I-J+1\le i \le I$ and $I-J+1\le i<m \le I$, respectively.
\end{theo}

This theorem generalizes W\"uthrich--Merz \cite[Theorem 2.15]{WM2015} to the exponential seniority weighting case, and we can also consider the non-informative prior limit $\gamma_j \downarrow 0$ to get the analogue to W\"uthrich--Merz \cite[formula (2.25)]{WM2015} for general $q_j \le 1$.

\section{Example}
\label{sec: Example}
We select the triangular data example for ${\cal D}_I$,  $I=J+1=10$, from  W\"uthrich--Merz \cite[Table 1.4]{WM2015}; the data are provided in Table \ref{CL 2} in the appendix below.
For this dataset, we compute Mack's \cite{Mack} CL reserves, Mack's conditional MSEP estimate, and the non-informative prior gamma-gamma Bayes' MSEP estimate; see W\"uthrich--Merz \cite[Table 2.4]{WM2015}.
We estimate the variance parameters $\sigma_j^2$ as in W\"uthrich--Merz \cite[Table 2.3]{WM2015}, and we keep these estimates fixed for all computations; the square roots $\widehat{\sigma}_j$ is given in Table \ref{CL 2} in the appendix.

In this example we consider $p_{i,j} \equiv 0$, $q_{i,j}\equiv q \in (0,1]$, and the non-informative prior limits $\gamma_j \to 0$.
The reserves are then for accident years $i+J> I$ given by
\begin{equation*}
 \widehat{R}^{{\rm CL}(q)}_i = \widehat{C}_{i,J}^{{\rm CL}(q)}-C_{i,d_i}
 = \lim_{\bgamma \downarrow \b0}\, \E_{\bgamma}\left[C_{i,J}\middle| {\cal D}_I\right]-C_{i,d_i},
\end{equation*}
and aggregated over all accident years, we obtain $\widehat{R}^{{\rm CL}(q)}_\bullet=\sum_{i=I-J+1}^I \widehat{R}^{{\rm CL}(q)}_i$. For $q=1$, we obtain Mack's \cite{Mack} CL reserves. 
The results are presented in Table \ref{CL MSEP results}.

\begin{table}[htb!]
\begin{center}
{\scriptsize
\begin{tabular}{|c||r|rrr|r|}
\hline
accident           & Mack's CL & Mack's&  Bayes'  & MC & MC error\\

year  $i$         &  reserves $\widehat{R}^{{\rm CL}(1)}_i$ &   RMSEP          &     RMSEP &
RMSEP     &  $\widehat{C}_i-\overline{C}_i$     \\
\hline\hline
        1 &          0  &            &            &          & 0 \\

        2 &     15,126 &        267 &        267 &      267 & 0  \\

        3 &     26,257 &        914 &        914 &      914 &-1\\

        4 &     34,538 &      3,058 &      3,058 &      3,056 &-5\\

        5 &     85,302 &      7,628 &      7,628 &      7,639 & 2\\

        6 &    156,494 &     33,341 &     33,341 &     33,333 &-38\\

        7 &    286,121 &     73,467 &     73,467 &     73,389 &-51\\

        8 &    449,167 &     85,398 &     85,399 &     85,347 &-89\\

        9 &  1,043,242 &    134,337 &    134,338 &     134,273 &-96\\

        10 &  3,950,815 &    410,817 &    410,850 &     411,816 &301\\
\hline
    total $\widehat{R}^{{\rm CL}(1)}_\bullet$ &  6,047,064 &    462,960 &    462,990 &      463,750 &21\\
\hline
\end{tabular}  
}
\caption{Mack's \cite{Mack} CL reserves and RMSEPs, non-informative prior
 gamma-gamma Bayes' CL RMSEPs of W\"uthrich--Merz \cite[Section 2.2]{WM2015},
 Monte Carlo RMSEP approximations from Algorithm \ref{algorithm conditional msep} and Monte Carlo
 errors $\widehat{C}_i-\overline{C}_i$ in reserves using one million Monte Carlo samples.}
\label{CL MSEP results}
\end{center}
\end{table}

The first two numerical columns of Table \ref{CL MSEP results} show Mack's \cite{Mack} CL reserves $\widehat{R}^{{\rm CL}(1)}_i$ and the resulting rooted mean squared error of prediction (RMSEP),
and the third column gives the RMSEP of the non-informative prior gamma-gamma Bayes' CL model. This latter model provides the identical CL reserves, and its RMSEP is lower bounded by Mack's RMSEP; see W\"uthrich--Merz \cite[Section 2.2.3]{WM2015}. The numbers in Table \ref{CL MSEP results} verify this lower bound property, as the third column dominates the second one.

For $q=1$, Algorithm \ref{algorithm conditional msep} should reproduce the non-informative gamma-gamma Bayes' CL results, up to a Monte Carlo simulation error. This is verified by the last two columns of Table \ref{CL MSEP results} which are based on one million Monte Carlo samples; note that the Monte Carlo errors can directly be compared to the reserves $\widehat{R}^{{\rm CL}(1)}_i$, because the payments $C_{i,d_i}$ are known at time $I$.

\begin{table}[htb!]
\begin{center}
{\scriptsize
\begin{tabular}{|c||rrr|rr|}
\hline
accident           & Mack's CL & Mack's&  Bayes'  & \multicolumn{2}{|c|}{exponential weighting $q=0.9$} \\

year  $i$         &  reserves $\widehat{R}^{{\rm CL}(1)}_i$ &   RMSEP          &     RMSEP &
reserves $\widehat{R}^{{\rm CL}(q)}_i$ & MC RMSEP    \\
\hline\hline
        1 &          0  &            &            &     0&      \\

        2 &     15,126 &        267 &        267 &   15,126&   274   \\

        3 &     26,257 &        914 &        914 &    26,283&  940 \\

        4 &     34,538 &      3,058 &      3,058 &     34,391& 3,139 \\

        5 &     85,302 &      7,628 &      7,628 &    84,594&  7,862 \\

        6 &    156,494 &     33,341 &     33,341 &    159,207 &34,357 \\

        7 &    286,121 &     73,467 &     73,467 &     289,417&75,679 \\

        8 &    449,167 &     85,398 &     85,399 &     457,342&88,209 \\

        9 &  1,043,242 &    134,337 &    134,338 &    1,035,100 &138,553\\

        10 &  3,950,815 &    410,817 &    410,850 &   3,888,217  &422,321 \\
\hline
    total $\widehat{R}^{{\rm CL}(q)}_\bullet$ &  6,047,064 &    462,960 &    462,990 &  5,989,679  &  482,923 \\
\hline
\end{tabular}  
}
\caption{Mack's \cite{Mack} CL reserves and RMSEPs, non-informative prior
  gamma-gamma Bayes' CL RMSEPs of W\"uthrich--Merz \cite[Section 2.2]{WM2015};
exponential seniority weighted CL reserves with $q=0.9$ and resulting RMSEP from Algorithm \ref{algorithm conditional msep} using one million Monte Carlo samples.}
\label{CL MSEP results exponential seniority}
\end{center}
\end{table}

The last two columns of Table \ref{CL MSEP results exponential seniority} give a non-informative prior example with constant exponential seniority weighting $q_{i,j}\equiv q=0.9$ (and $p_{i,j}\equiv 0$), i.e., this applies an exponential seniority weighting as illustrated in Figure \ref{Figure4} (rhs). The resulting reserves $\widehat{R}^{{\rm CL}(q)}_\bullet$ are smaller than Mack's CL reserves $\widehat{R}^{{\rm CL}(1)}_\bullet$. This comes from the fact that the observations $C_{i,j+1}/C_{i,j}$ have a slightly decreasing trend in the left-lower triangle of the observations ${\cal D}_I$, see Table \ref{individual observations}, which leads to smaller CL factor estimates through the exponential seniority weighting.

\begin{table}[htbp]
  \centering
  \small
  \setlength{\tabcolsep}{5pt}
  \renewcommand{\arraystretch}{1.15}
{\scriptsize
  \begin{tabular}{r|r*{9}{r}}
    \toprule
$i/j$     & 0 & 1 & 2 & 3 & 4 & 5 & 6 & 7 & 8 & 9 \\
    \midrule
    1 & 1.6257 & 1.0926 & 1.0197 & 1.0192 & 1.0057 & 1.0060 & 1.0013 & 1.0010 & 1.0014 & ~\qquad~ \\
    2 & 1.5115 & 1.0754 & 1.0147 & 1.0065 & 1.0035 & 1.0050 & 1.0011 & 1.0011 &  &  \\
    3 & 1.4747 & 1.0916 & 1.0260 & 1.0147 & 1.0062 & 1.0051 & 1.0008 &  &  &  \\
    4 & 1.4577 & 1.0845 & 1.0206 & 1.0141 & 1.0092 & 1.0045 &  &  &  &  \\
    5 & 1.4750 & 1.0767 & 1.0298 & 1.0244 & 1.0109 &  &  &  &  &  \\
    6 & 1.4573 & 1.0635 & 1.0255 & 1.0107 &  &  &  &  &  &  \\
    7 & 1.5166 & 1.0663 & 1.0249 &  &  &  &  &  &  &  \\
    8 & 1.4614 & 1.0683 &  &  &  &  &  &  &  &  \\
    9 & 1.4457 &  &  &  &  &  &  &  &  &  \\
    10 &  &  &  &  &  &  &  &  &  &  \\
    \bottomrule
  \end{tabular}}
\caption{Individual observations $C_{i,j+1}/C_{i,j}$ in ${\cal D}_I$.}
\label{individual observations}
\end{table}

\begin{figure}[htb!]
  \begin{center}
    \begin{minipage}[t]{0.475\textwidth}
\begin{center}
\includegraphics[width=\textwidth]{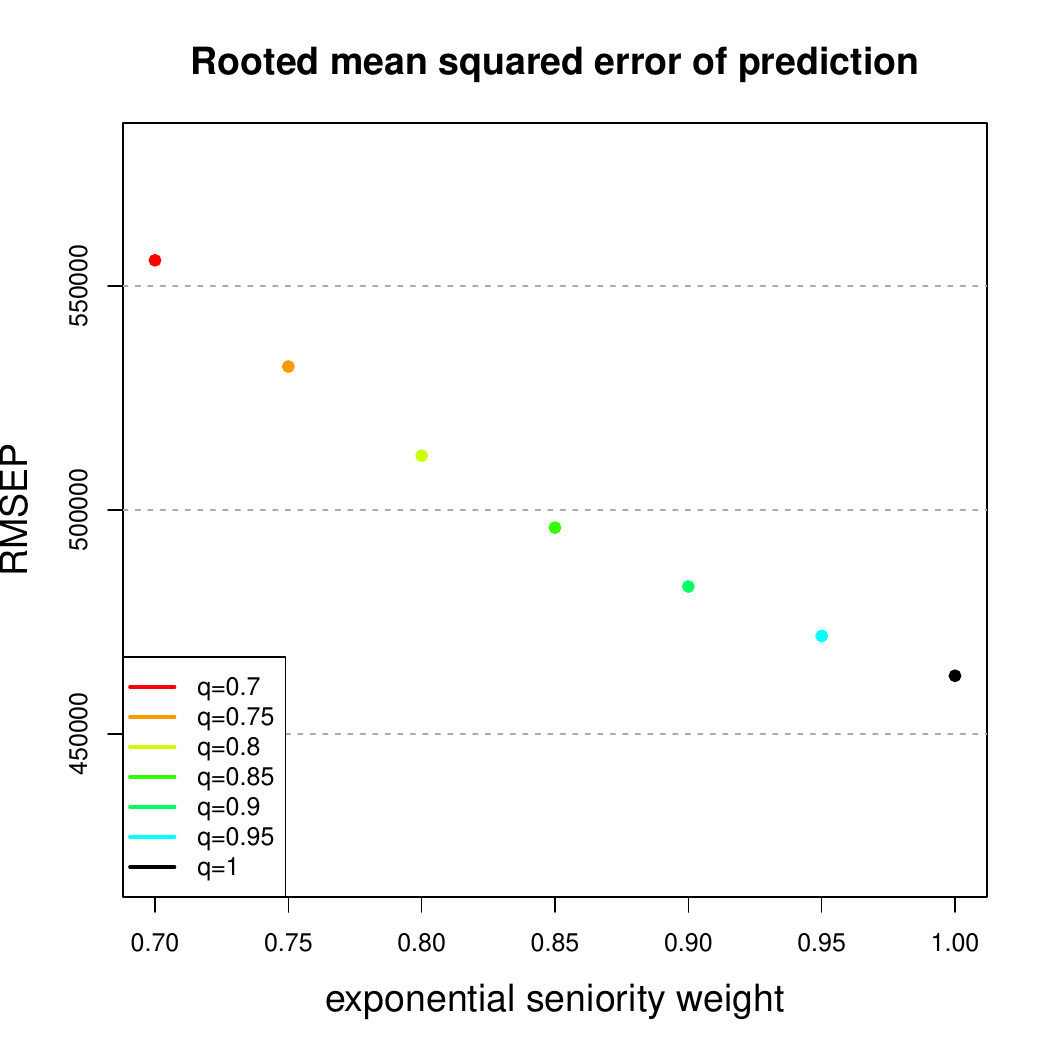}
\end{center}
\end{minipage}
\begin{minipage}[t]{0.475\textwidth}
\begin{center}
\includegraphics[width=\textwidth]{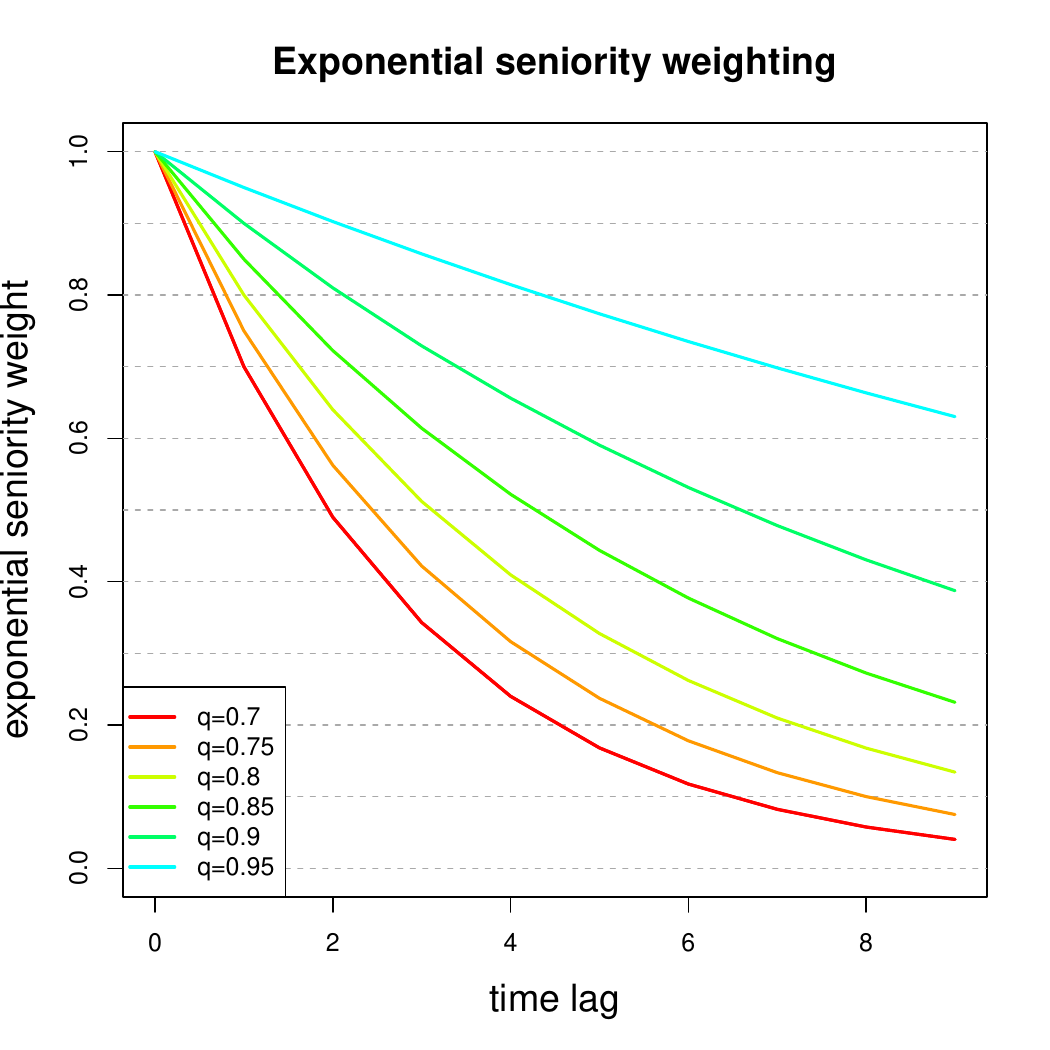}
\end{center}
\end{minipage}
\end{center}
\vspace{-.7cm}
\caption{(lhs) Total RMSEPs as a function of $q\in \{0.70, 0.75, \ldots, 1.00\}$, (rhs) exponential seniority weights $q^k$, for $k\in \{0,\ldots, 9\}$ and $q\in \{0.70, 0.75, \ldots, 0.95\}$.}
\label{Figure5}
\end{figure}

Interestingly, the conditional RMSEPs in the last column of Table \ref{CL MSEP results exponential seniority} increase for $q=0.9$ compared to the CL case $q=1$. Intuitively, this is explained by the fact that an exponential seniority weighting assigns less weight to older claims, hence, relying more on the most recent claims. This can be interpreted as a decrease of the amount of past information used, which leads to a bigger uncertainty in the CL factor estimation. For small $q$, we almost exclusively rely on the latest observed diagonal $i+j=I$, i.e., in that case, the CL factor $\widehat{f}_j^{(I)}$ is estimated almost entirely from a single observation $C_{r_{j}-1,j+1}/C_{r_{j}-1,j}$, see \eqref{CL factor for examples}. Figure \ref{Figure5} (rhs) shows the resulting exponential seniority weights using $q\in \{0.70, 0.75, \ldots, 0.95\}$. We observe that for $q=0.70$, observations with a time lag of more than 4 periods to the last observed diagonal receive weights of less than 20\%.

Figure \ref{Figure5} (lhs) shows the total RMSEPs for different seniority weighting 
$q\in \{0.70, 0.75, \ldots, 1.00\}$: the smaller the weight $q$, the less weight is assigned to older observations, and the bigger the RMSEP becomes, because older observations are less informative. If we use the interpretation
of the gamma-thinning construction presented in Ahn et al.~\cite{Ahn1}, this is justified by the fact that the state-space update is thinning past information and it is adding new noise to the latent (memory) process $\Theta_{1:I,j}$.

\medskip

In view of the above results, there is the question of an optimal choice of the parameter $q\in (0,1]$. So far, it has been treated as a hyper-parameter selected by the modeler. An elegant approach is the re-reserving and model validation method by Balona--Richman \cite{BalonaRichman}. The idea is to work through the triangles/trapezoids 
\begin{equation*}
  {\cal D}_t = \sigma\left\{ C_{i,j};~i+j\le t, \, 1\le i \le I, \, 0 \le j \le J \right\} \qquad \text{ for $t \in \{1,\ldots, I-1\}$},
\end{equation*}
by making predictions based on ${\cal D}_t$ and validating them against ${\cal D}_{t+1}$, that is, by successively adding new diagonals to the prediction and validation process (note that there are not necessarily observations for all development periods $j$). This can be done for different values of $q$, and the one that has the best validation performance is the one selected for extrapolation beyond time $I$. In the same way, the band in Figure \ref{Figure1} (rhs) can be selected. Note that this is an out-of-sample validation, because one validates against new observations in ${\cal D}_{t+1}\setminus {\cal D}_t$ and the resulting predictions at time $t+1$.

Having an analytical full data likelihood, see Corollary \ref{data likelihood lemma}, one can alternatively perform maximum likelihood estimation (MLE). To keep the estimation complexity under control, we use the CL estimates for the variance parameters $\sigma_j^2$, given in Table \ref{CL 2}. This is an ad-hoc solution because these parameter estimates are based on the choice $q=1$, and alternatively one could employ an iterative estimation procedure. For simplicity, we do not do this here. Thus, we fix $\sigma_j^2=\widehat{\sigma}_j^2$, we set $p_{i,j} \equiv 0$, we consider the non-informative prior limit $\gamma_j \downarrow 0$, and we optimize the resulting data log-likelihood over
$q_{i,j} \equiv q \in (0,1]$. It is given by
\begin{equation}\label{log-likelihood}
q ~ \mapsto ~\ell_{{\cal D}_I}(q)
    = \sum_{j=1}^{J} \sum_{i=2}^{r_j}
\log K \left(C_{i,j}\middle| \, C_{i,j-1}/\widehat{\sigma}^2_{j-1}, \, a_{i,j-1}^-, \, \widehat{\sigma}^2_{j-1}b_{i,j-1}^-\right),
\end{equation}
with $a_{2,j}^-=q C_{1,j}/\widehat{\sigma}^2_{j}$ and $b_{2,j}^-=q C_{1,j+1}/\widehat{\sigma}^2_{j}$, and further updates, see \eqref{recursion formula 1}-\eqref{recursion formula a plus},
\begin{equation*}
a_{i+1,j}^-=q\left(a_{i,j}^-+{C_{i,j}}/{\widehat{\sigma}_j^2}\right) \qquad \text{ and } \qquad
  b_{i+1,j}^-= q\left(b_{i,j}^-+{C_{i,j+1}}/{\widehat{\sigma}_j^2}\right).
\end{equation*}                 
This corresponds to the non-informative prior limit $\gamma_j \downarrow 0$, which requires us to drop accident year $i=1$ from the log-likelihood \eqref{log-likelihood}.


Numerical maximization provides $\widehat{q}^{\rm MLE}=0.717$, with a log-likelihood value of $-447.36$, the choice $q=0.9$ gives a log-likelihood value of $-448.05$, and $q=1.0$ a value of $-448.87$. Thus, we should select a $q$-value that is close to the red curve $q=0.70$ in Figure \ref{Figure5} (rhs).

\medskip

\begin{table}[htb!]
 \begin{center}
 {\scriptsize
\begin{tabular}{|c||rrr|rr|}
\hline
accident           & Mack's CL & MW \cite{MW2008}&  Bayes'  & \multicolumn{2}{|c|}{exponential weighting $q=0.9$} \\

year  $i$         &  reserves $\widehat{R}^{{\rm CL}(1)}_i$ &   CDR RMSEP          &   CDR  RMSEP &
reserves $\widehat{R}^{{\rm CL}(q)}_i$ & CDR RMSEP    \\
\hline\hline
        1 &          0  &            &            &     0&      \\

        2 &     15,126 &        267 &        267 &   15,126&   274   \\

        3 &     26,257 &        884 &        884 &    26,283&  910 \\

        4 &     34,538 &      2,948 &      2,948 &     34,391& 3,029 \\

        5 &     85,302 &      7,018 &      7,018 &    84,594&  7,227 \\

        6 &    156,494 &     32,470 &     32,470 &    159,207 &33,448 \\

        7 &    286,121 &     66,178 &     66,178 &     289,417&68,223 \\

        8 &    449,167 &     50,296 &     50,296 &     457,342&52,400 \\

        9 &  1,043,242 &    104,311 &    104,311 &    1,035,100 &107,618\\

        10 &  3,950,815 &    385,773 &    385,804 &   3,888,217  &395,735 \\
\hline
    total $\widehat{R}^{{\rm CL}(q)}_\bullet$ &  6,047,064 &    420,220 &    420,249 &  5,989,679  &  436,004 \\
\hline
\end{tabular}  
}
\caption{Claims reserves and CDR RMSEP: Merz--W\"uthrich (MW) formula \cite{MW2008},
  gamma-gamma Bayes' CL CDR uncertainty W\"uthrich--Merz \cite[Theorem 2.15]{WM2015}, this corresponds to $q=1$,
  and CDR RMSEPs for $q=0.9$ using Theorem \ref{theo WM CDR}.}
\label{CL MSEP results CDR}
\end{center}
\end{table}

Next, we turn our attention to the one-year CDR analysis. The results are presented in Table \ref{CL MSEP results CDR}, and they are essentially the same as for the total RMSEP. Namely, the Bayes' version with $q=1$ dominates the 
Merz--W\"uthrich (MW) CDR formula \cite{MW2008}, decreasing $q$ increases the CDR RMSEP, and generally, we have
\begin{equation*}
  \operatorname{msep}_{C_\bullet|{\cal D}_I}\left(\widehat{C}_\bullet\right) \ge
  \operatorname{msep}_{\operatorname{CDR}|{\cal D}_I}\left(0\right),
\end{equation*}
which follows from a martingale argument; see W\"uthrich--Merz \cite[Corollary 1.2]{WM2015}.
The gap between the two RMSEPs is relatively small in our example because the data shows a fast run-off, see Table \ref{CL 2}. This concludes the example.

\section{Conclusion}
\label{sec: Conclusion}

We presented an observation-driven state-space version of the gamma-gamma Bayes' CL model of W\"uthrich--Merz \cite{WM2015}. This observation-driven state-space dynamics considers a gamma latent stochastic process as proposed in Ahn et al.~\cite{Ahn2}. This leads to a full analytical representation of the data likelihood, which allows us to compute the resulting claims reserves, the conditional mean squared error of prediction of the ultimate claims can be determined by a simple Monte Carlo simulation, and the one-year claims development result uncertainty is available in closed form.

This observation-driven gamma-gamma state-space model has the properties that it can replicate the chain-ladder reserves under non-informative priors and under specific hyper-parameter choices. These hyper-parameters also admit an exponential seniority weighting in past information entering the chain-ladder computation, and the tractable data likelihood even allows one to estimate the  optimal seniority weights with maximum likelihood estimation.

To the best of our knowledge, this is the first fully tractable model that allows for seniority weighting and at the same time it obeys the crucial assumptions of Mack's \cite{Mack} distribution-free chain-ladder model, in particular, the proportional linear conditional variance behavior.

There are similar results on the chain-ladder factor estimation side, see Marais--Grove \cite{MaraisGrove}, but these results are not obtained in a full stochastic reserving model, they are rather enforced manually. Therefore, they do not allow for a consistent computation of the prediction uncertainty. All our results are a natural consequence of the stated model assumptions.

\paragraph{AI declaration.} The author developed the concept of the paper, all mathematical results, and he has fully written the first version of the paper. ChatGPT-6 was used to improve the manuscript, to check the mathematical results and to implement the examples.

\medskip

{\small 
\renewcommand{\baselinestretch}{.51}
}

\newpage

\appendix

\section{Proofs}
In this appendix, we give all mathematical proofs.

\medskip

{\Beweis {\bf Proof of Lemma \ref{cor mean-stationarity}.} We have ${\cal B}_j \subset {\cal H}_{i,j}$, in particular, all cumulative claims $C_{1:I,j}$ are given to model the next ones $C_{i,j+1}$ for $i\in \{1,\ldots, I\}$.
  From \eqref{credibility 2}, we obtain for $i\in \{1,\ldots, I-1\}$, since $\omega_{i,j}$ are ${\cal B}_j$-measurable,
  \begin{eqnarray*}
\E \left[ \Theta^{-1}_{i+1,j}\middle| {\cal B}_j\right]
    &=&
\Delta_{i,j}\left(\omega_{i,j}\, \E \left[ \frac{C_{i,j+1}}{C_{i,j}}\middle| {\cal B}_j\right] + \left(1-\omega_{i,j}\right)\E \left[f^-_{i,j}\middle| {\cal B}_j\right]\right)
        +\left(1-\Delta_{i,j}\right) f_j
\\ &=&
\Delta_{i,j}\left(\omega_{i,j}\, \E \left[\frac{1}{C_{i,j}}\, \E \left[ C_{i,j+1}\middle| {\cal H}_{i-1,j}, {\cal G}_{i,j}\right]  \middle| {\cal B}_j\right] + \left(1-\omega_{i,j}\right)\E \left[\Theta^{-1}_{i,j}\middle| {\cal B}_j\right]\right)
       +\left(1-\Delta_{i,j}\right) f_j\\ &=&
\Delta_{i,j}\E \left[\Theta^{-1}_{i,j}\middle| {\cal B}_j\right]
+\left(1-\Delta_{i,j}\right) f_j.                      
\end{eqnarray*}
The proof then follows by induction because
$\E [\Theta^{-1}_{1,j}| {\cal B}_j]  =\E [\Theta^{-1}_{1,j}| {\cal H}_{0,j}]  =f_j$.
\EndProof}

\bigskip

{\Beweis {\bf Proof of Lemma \ref{likelihood lemma}.}
Select $j \in \{0,\ldots, J-1\}$.
We begin by considering the density of $C_{1:I,j+1}$, given ${\cal B}_j$,
\begin{equation}\label{factorization of density}
  f\left(C_{1:I,j+1} \middle| {\cal B}_j\right)
  = f\left(C_{I,j+1} \middle| {\cal H}_{I-1,j}\right)f\left(C_{1:I-1,j+1} \middle| {\cal B}_j\right)
  = \prod_{i=1}^I f\left(C_{i,j+1} \middle| {\cal H}_{i-1,j}\right).
\end{equation}
Next, we compute each term in the product of \eqref{factorization of density}
\begin{eqnarray*}
  f\left(C_{i,j+1} \middle| {\cal H}_{i-1,j}\right)
  &=& \int_0^\infty
  f\left(C_{i,j+1} \middle| {\cal H}_{i-1,j}, \Theta_{i,j}\right)
      f\left(\Theta_{i,j} \middle| {\cal H}_{i-1,j}\right) {\rm d}\Theta_{i,j}
 \\ &=&\int_0^\infty \frac{(\Theta_{i,j}/\sigma_{j}^2)^{C_{i,j}/\sigma_{j}^2}}{\Gamma(C_{i,j}/\sigma_{j}^2)}
        C_{i,j+1}^{C_{i,j}/\sigma_{j}^2-1}e^{-\Theta_{i,j}/\sigma_{j}^2 C_{i,j+1}}
        \frac{(b_{i,j}^-)^{1+a_{i,j}^-}}{\Gamma(1+a_{i,j}^-)} \Theta_{i,j}^{a_{i,j}^-} e^{-b_{i,j}^- \Theta_{i,j}}
        {\rm d}\Theta_{i,j}
 \\ &=& K \left(C_{i,j+1}\middle| C_{i,j}/\sigma_j^2, a_{i,j}^-,\sigma_j^2 b_{i,j}^-\right).       
\end{eqnarray*}
The development year factorization follows from the definition.
This completes the proof. 
\EndProof}

\bigskip

{\Beweis {\bf Proof of Proposition \ref{Chain Ladder Factor Proposition}.}
We need to align the information set ${\cal D}_I$ and 
${\cal H}_{r_j-1,j}$, see Figure \ref{Figure3}. We start from 
Corollary \ref{data likelihood lemma}. The information set
${\cal H}_{r_j-1,j}$ contains the observations $C_{1:r_{j}-1,j+1}=
C_{1:r_{j+1},j+1}$, this is the yellow column in Figure \ref{Figure3} (rhs). We condition on these observations in the data likelihood
\begin{eqnarray*}
    f\left((C_{i,k})_{i+k \le I,\, k\ge 1, \, k \neq j+1} \middle| C_{1:I,0}, C_{1:r_{j+1},j+1}\right)
    &\propto& \prod_{k=1}^{j+1} \prod_{i=1}^{r_k}
        K \left(C_{i,k}\middle| \, C_{i,k-1}/\sigma^2_{k-1}, \, a_{i,k-1}^-, \, \sigma^2_{k-1}b_{i,k-1}^-\right)
\\&&\times~ \prod_{k=j+2}^{J} \prod_{i=1}^{r_k}
        K \left(C_{i,k}\middle| \, C_{i,k-1}/\sigma^2_{k-1}, \, a_{i,k-1}^-, \, \sigma^2_{k-1}b_{i,k-1}^-\right).
   \end{eqnarray*}
Thus, conditioning on $C_{1:r_{j+1},j+1}$ leads to an independent decoupling of ${\cal B}_j$ and the upper right trapezoid 
${\cal D}_{I; j+2:J}=(C_{i,k})_{i+k \le I,\,  j+2 \le  k \le J}$. The same argument holds for ${\cal B}_j \vee \sigma(C_{r_{j+1}+1:I, j+1})$
being conditionally independent of the upper right trapezoid ${\cal D}_{I; j+2:J}$, given $C_{1:r_{j+1},j+1}$. Therefore, this upper right trapezoid does not impact the computation of $f_{i,j}^-$, given ${\cal H}_{r_{j+1},j}$. This conditional independent decoupling implies
for $i \ge r_j=r_{j+1}+1$
\begin{eqnarray*}
\E\left[f_{i,j}^- \middle| {\cal D}_{I}\right]
&=& \E\left[\E\left[f_{i,j}^- \middle| {\cal H}_{r_{j+1},j}, {\cal D}_{I}\right] \middle| {\cal D}_{I}\right]
~=~
\E\left[\E\left[f_{i,j}^- \middle| {\cal H}_{r_{j+1},j}\right] \middle| {\cal D}_{I}\right]
\\
  &=& 
\E\left[\left(\prod_{k=r_{j+1}+1}^{i-1} \Delta_{k,j}\right) f_{r_{j+1}+1,j}^-
    + \left(1-\prod_{k=r_{j+1}+1}^{i-1} \Delta_{k,j}\right) f_j \middle| {\cal D}_{I}\right],
\end{eqnarray*}
the last step uses Corollary \ref{corollary unbiasedness posterior}.
The claim then follows because $r_{j+1}+1=r_j$ and because
all terms in the last expression are ${\cal D}_I$-measurable, i.e., 
in particular, $f_{r_{j},j}^-$.
\EndProof}

\bigskip

{\Beweis
{\bf Proof of Theorem \ref{CL forecast}.}
We use the conditional independent property used in
Proposition \ref{Chain Ladder Factor Proposition} to operate with information sets ${\cal H}_{i,j}$, $i\ge r_{j}$, and dropping the upper right trapezoid of ${\cal D}_I$ for the development periods $j+2,\ldots, J$. 
For $i\ge I-J+1=r_{J-1}$, we first apply the CL step
\begin{equation*}
  \widehat{C}_{i,J}
  = \E \left[C_{i,J} \middle| {\cal D}_I \right]
 =\E \left[\E \left[C_{i,J} \middle|{\cal H}_{i-1,J-1}, {\cal D}_I \right] \middle| {\cal D}_I \right]     
= \E \left[C_{i,J-1}\, f_{i,J-1}^- \middle| {\cal D}_I \right]. 
\end{equation*}
In the second step, we compute the ${\cal D}_I$-measurable CL factor
\begin{eqnarray*}
  \widehat{C}_{i,J}
  &=& \E \left[C_{i,J} \middle| {\cal D}_I \right]~=~ \E \left[C_{i,J-1}\, f_{i,J-1}^- \middle| {\cal D}_I \right]
  \\&=& \E \left[\E \left[C_{i,J-1}\, f_{i,J-1}^- \middle|{\cal H}_{r_{J-1}-1,J-1}, {\cal D}_I \right] \middle| {\cal D}_I \right]
  \\&=&\E \left[C_{i,J-1}\,  \E  \left[ f_{i,J-1}^- \middle|{\cal H}_{r_{J-1}-1,J-1}, {\cal D}_I \right] \middle| {\cal D}_I \right]
~=~ \widehat{f}^{(I)}_{i,J-1}\,
\E \left[C_{i,J-1} \middle| {\cal D}_I \right]        ,     
\end{eqnarray*}
the last step used Proposition \ref{Chain Ladder Factor Proposition}.
This is recursively iterated until development period $j=d_i$.
\EndProof}

\bigskip

{\Beweis {\bf Proof of Theorem \ref{CDR theorem}.}
From \eqref{simplifying CDR} and \eqref{new one-year representation}, we have  
\begin{eqnarray*}
  &&\hspace{-1cm}
     \operatorname{msep}_{\operatorname{CDR}|{\cal D}_I}\left(0\right)~=~
  {\rm Var}\left(  \sum_{i=1}^I \E \left[C_{i,J} \middle| {\cal D}_{I+1} \right]\middle| {\cal D}_I\right)
  \\&=&
      {\rm Var}\left(  \sum_{i=I-J+1}^I \prod_{j=0}^{J-1} \left(\nu_{i,j} + \kappa_{i,j}C_{r_j,j+1}\right)\middle| {\cal D}_I\right)
  \\&=& \sum_{i=I-J+1}^I \sum_{k=I-J+1}^I {\rm Cov}\left(   \prod_{j=0}^{J-1} \left(\nu_{i,j} + \kappa_{i,j}C_{r_j,j+1}\right), \prod_{j=0}^{J-1} \left(\nu_{k,j} + \kappa_{k,j}C_{r_j,j+1}\right)\middle| {\cal D}_I\right).   
\end{eqnarray*}
Using the standard representation of covariances and the conditionally independent decoupling \eqref{conditional independence} along the diagonal yields
\begin{eqnarray*}
  &&\hspace{-1cm}{\rm Cov}\left(   \prod_{j=0}^{J-1} \left(\nu_{i,j} + \kappa_{i,j}C_{r_j,j+1}\right), \prod_{j=0}^{J-1} \left(\nu_{k,j} + \kappa_{k,j}C_{r_j,j+1}\right)\middle| {\cal D}_I\right)
  \\&=&\E\left[   \prod_{j=0}^{J-1} \left(\nu_{i,j} + \kappa_{i,j}C_{r_j,j+1}\right)  \left(\nu_{k,j} + \kappa_{k,j}C_{r_j,j+1}\right)\middle| {\cal D}_I\right]-\widehat{C}_{i,J}\widehat{C}_{k,J}
  \\&=&\prod_{j=0}^{J-1}\E\left[    \left(\nu_{i,j} + \kappa_{i,j}C_{r_j,j+1}\right)  \left(\nu_{k,j} + \kappa_{k,j}C_{r_j,j+1}\right)\middle| {\cal D}_I\right]-\widehat{C}_{i,J}\widehat{C}_{k,J}
  \\&=&\prod_{j=0}^{J-1}\left[\nu_{i,j}\nu_{k,j}+(\nu_{i,j}\kappa_{k,j}
+\nu_{k,j}\kappa_{i,j})\mu_{j} + \kappa_{i,j} \kappa_{k,j}\tau_j^2\right]-\widehat{C}_{i,J}\widehat{C}_{k,J}.                
\end{eqnarray*}
This completes the proof.
\EndProof}

\bigskip

{\Beweis {\bf Proof of Theorem \ref{theo WM CDR}.}
  Introduce the normalized quantities
  \begin{equation*}
    Z_{j}= \frac{C_{r_j,j+1}}{\widehat{f}_j^{(I)} C_{r_j,j}} \qquad \text{ and } \qquad
    H_j = 1 + \omega_j^\star \left(Z_j-1\right).
  \end{equation*}
These are precisely the variables that contribute to the randomness in \eqref{new one-year representation}.  
We have conditional means
\begin{equation*}
  \E \left[ Z_j \middle| {\cal D}_I\right] =   \E \left[ H_j \middle| {\cal D}_I\right]=1,
\end{equation*}
this follows from \eqref{eq 11}. Using \eqref{eq 22} we have
\begin{equation*}
  \operatorname{Var}\left( Z_j \middle| {\cal D}_I\right)
  = \frac{\operatorname{Var}\left( C_{r_j,j+1} \middle| {\cal D}_I\right)}{(\widehat{f}_j^{(I)} C_{r_j,j})^2}
  = \frac{C_{r_j,j} + \sigma_j^2 A_j}{(A_j-1)C_{r_j,j}}
  = \frac{\Psi_j}{\omega_j^\star}.
  \end{equation*}
  This implies
  \begin{equation}\label{crucial terms CDR proof}
    \E \left[ Z_j^2 \middle| {\cal D}_I\right]=1+\frac{\Psi_j}{\omega_j^\star},
    \qquad
    \E \left[ H_j^2 \middle| {\cal D}_I\right] = 1+ \omega_j^\star\Psi_j
    \qquad \text{ and } \qquad
    \E \left[ Z_jH_j \middle| {\cal D}_I\right] = 1+ \Psi_j.
  \end{equation}
  Observe that we can write the one-year update of the CL factor as
  \begin{equation*}
    \widehat{f}^{(I+1)}_{j} = \omega_j^\star \,\frac{C_{r_j,j+1}}{C_{r_j,j}}+ \left(1-\omega_j^\star\right) \widehat{f}^{(I)}_{j}
    = \widehat{f}^{(I)}_{j} \left( \omega_j^\star \,Z_j+ \left(1-\omega_j^\star\right) \right)
    = \widehat{f}^{(I)}_{j} H_j.
  \end{equation*}
  Thus, we can rewrite \eqref{new one-year representation} as
\begin{equation*}
\E \left[C_{i,J} \middle| {\cal D}_{I+1} \right]
=C_{i,d_i+1}\, \prod_{j=d_i+1}^{J-1}
\widehat{f}^{(I+1)}_{i,j}
=\widehat{C}_{i,J}\left(Z_{d_i} \prod_{j=d_i+1}^{J-1} H_j\right).
\end{equation*}
The claim for the individual accident years $i$ now immediately follows from the conditionally independent decoupling \eqref{conditional independence}, given ${\cal D}_I$, and from \eqref{crucial terms CDR proof}.

The aggregated version follows completely analogously by considering the conditional
covariances for $i<m$, which gives $d_i>d_m$,
\begin{equation*}
  \operatorname{Cov}\left(\operatorname{CDR}_i,\operatorname{CDR}_m\middle| {\cal D}_I\right)
  =\widehat{C}_{i,J}\,\widehat{C}_{m,J}\left(Z_{d_i} \prod_{j=d_i+1}^{J-1} H_j\right)\left(Z_{d_m} \prod_{j=d_m+1}^{J-1} H_j\right),
\end{equation*}
with the difference that the term $Z_{d_i}^2$ needs to be replaced by $Z_{d_i}H_{d_i}$.
This completes the proof.
\EndProof}

\begin{landscape}
\begin{table}[ht]
\begin{center}
{\scriptsize
\begin{tabular}{|c||rrrrrrrrrr|}
\hline
accident& \multicolumn{10}{|c|}{development year $j$}\\
year $i$           &          0 &          1 &          2 &          3 &          4 &          5 &          6 &          7 &          8 &          9 \\
\hline\hline
         1 &  5,946,975 &  9,668,212 & 
10,563,929 & 10,771,690 & 10,978,394 & 11,040,518 & 11,106,331 & 11,121,181 & 11,132,310 & 11,148,124 \\

         2 &  6,346,756 &  9,593,162 & 10,316,383 & 
10,468,180 & 10,536,004 & 10,572,608 & 10,625,360 & 10,636,546 & 10,648,192 &            \\

         3 &  6,269,090 &  9,245,313 & 10,092,366 & 
10,355,134 & 10,507,837 & 10,573,282 & 10,626,827 & 10,635,751 &            &            \\

         4 &  5,863,015 &  8,546,239 &  9,268,771 &  9,459,424 &  9,592,399 &  9,680,740 &  9,724,068 &            &            &            \\

         5 &  5,778,885 &  8,524,114 &  9,178,009 &  9,451,404 &  9,681,692 &  9,786,916 &            &            &            &            \\

         6 &  6,184,793 &  9,013,132 &  9,585,897 &  9,830,796 &  9,935,753 &            &            &            &            &            \\

         7 &  5,600,184 &  8,493,391 &  9,056,505 &  9,282,022 &            &            &            &            &            &           \\

         8 &  5,288,066 &  7,728,169 &  8,256,211 &            &            &            &            &            &            &            \\

         9 &  5,290,793 &  7,648,729 &            &            &            &            &            &            &            &            \\

         10 &  5,675,568 &            &            &            &            &            &            &            &            &            \\
\hline
 $\widehat{f}_j^{(I)}$ &     1.4925 &     1.0778 &      1.0229 &     1.0148 &     1.0070 &     1.0051 &     1.0011 &     1.0010 &     1.0014 &            \\
$\sqrt{\widehat{\sigma}^2_j}$&
90.62&31.36&15.41&19.56&9.27&1.99&0.82&0.22&0.06&\\
  \hline
\end{tabular}  
}

  \caption{Observed cumulative claims $C_{i,j}$, $i+j\le I$, 
the classical Mack's CL factors $\widehat{f}^{(I)}_j$ given in \eqref{classical Mack CL factors}, and the (rooted) variance estimates taken from W\"uthrich--Merz \cite[Table 2.3]{WM2015}.}
\label{CL 2}
\end{center}
\end{table}
\end{landscape}

\end{document}